\documentclass[12pt,fleqn]{article}
\usepackage{amsfonts,ulem,url}
\usepackage{amssymb}
\usepackage{amsmath}
\usepackage{amstext,verbatim,graphicx,ifthen,multirow,amsthm}
\usepackage{bm}
\usepackage{natbib}
\usepackage[labelfont=bf]{caption}
\usepackage{subcaption}
\usepackage{setspace}
\usepackage[final]{pdfpages}
\usepackage{epstopdf}
\usepackage{mathtools}
\usepackage{tensor}
\usepackage[hyperindex,breaklinks]{hyperref}

\renewcommand{\mathbf}{\boldsymbol}
\renewcommand{\thepage}{}
\renewcommand{\appendix}{\footnotesize\parindent 0cm\setcounter{equation}{0}
\renewcommand{\theequation}{A.\arabic{equation}}
\setcounter{lemma}{0}\renewcommand{\thelemma}{A.\arabic{lemma}}}

\newcommand{\opt}{{\rm opt}}

\newcommand{\elastic}{{\rm en}}

\newcommand{\did}{{\rm did}}
\newcommand{\adh}{{\rm adh}}
\newcommand{\constr}{{\rm constr}}
\newcommand{\subs}{{\rm subset}}
\newcommand{\single}{{\rm single}}

\newcommand{\bz}{\mathbf{Y}}

\def\monthname{\ifcase\month\or
  January\or February\or March\or April\or May\or June\or July\or
  August\or September\or October\or November\or December\fi}
\numberwithin{equation}{section}

\newcommand{\obs}{{\rm obs}}

\newcommand{\bge}{\begin{equation}}
\newcommand{\ene}{\end{equation}}
\newcommand{\by}{{\mathbf{Y}^\obs}}
\newcommand{\byzn}{\mathbf{Y}^\obs_{{\rm c},{\rm pre}}}
\newcommand{\byze}{\mathbf{Y}^\obs_{{\rm t},{\rm pre}}}
\newcommand{\byen}{\mathbf{Y}^\obs_{{\rm c},{\rm post}}}
\newcommand{\byee}{\mathbf{Y}^\obs_{{\rm t},{\rm post}}}

\newcommand{\bnyzn}{\mathbf{Y}_{{\rm c},{\rm pre}}(0)}
\newcommand{\bnyze}{\mathbf{Y}_{{\rm t},{\rm pre}}(0)}
\newcommand{\bnyen}{\mathbf{Y}_{{\rm c},{\rm post}}(0)}
\newcommand{\bnyee}{\mathbf{Y}_{{\rm t},{\rm post}}(0)}
\newcommand{\btyee}{\mathbf{Y}_{{\rm t},{\rm post}}(1)}

\newcommand{\bx}{{\mathbf{X}}}
\newcommand{\bxn}{\bx_{\rm c}}
\newcommand{\bxe}{\bx_{\rm t}}

\newcommand{\ols}{{\rm ols}}

\newcommand{\mmv}{{\mathbb{V}}}

\DeclareMathOperator*{\argmin}{arg\,min}

\def\monthname{\ifcase\month\or
January\or February\or March\or April\or May\or June\or
July\or August\or September\or October\or November\or December\fi}
\renewcommand{\appendix}{\small\parindent 0cm\setcounter{equation}{0}
\renewcommand{\theequation}{A.\arabic{equation}}
\setcounter{lemma}{0}\renewcommand{\thelemma}{A.\arabic{lemma}}
\setcounter{theorem}{0}\renewcommand{\thetheorem}{A.\arabic{theorem}}}
\begin{document}

\title{\textbf{Balancing, Regression, Difference-In-Differences and Synthetic Control Methods: A Synthesis}\thanks{{\small We are
grateful for comments by seminar participants at the California Econometrics Conference and Tim Squires.}} }
\author{Nikolay Doudchenko\thanks{{\small  Graduate School of Business, Stanford University, nikolayd@stanford.edu.}} \and Guido W. Imbens\thanks{{\small Professor of
Economics,
Graduate School of Business, Stanford University, SIEPR, and NBER,
imbens@stanford.edu.}} }
\date{ Current version \ifcase\month\or
January\or February\or March\or April\or May\or June\or
July\or August\or September\or October\or November\or December\fi \ \number%
\year\ \  }
\maketitle\thispagestyle{empty}

\begin{abstract}
\singlespacing

 In a seminal paper \cite*{abadie2010} (ADH), see also \citet{abadie2003, abadie2014}, develop the synthetic control procedure for estimating the effect of a treatment, in the presence of a single treated unit and a number of control units, with pre-treatment outcomes observed for all units. The method constructs a set of weights such that selected  covariates and  pre-treatment outcomes of the treated unit are approximately matched by a weighted average of control units (the synthetic control).
The weights are  restricted to be nonnegative and  sum  to one, which is important because it allows the procedure to obtain unique weights even when the number of lagged outcomes is modest relative to the number of control units, a common setting in applications.  In the current paper we propose a  generalization that allows  the weights to be negative, and their sum to differ from one, and that allows for a permanent additive difference between the treated unit and the controls, similar to  difference-in-difference procedures. The weights directly minimize the distance between the lagged outcomes for the treated and the control units, using regularization methods to deal with a potentially large number of possible control units. 
\end{abstract}

\noindent \textbf{Keywords}: comparative study, synthetic control, difference-in-differences, matching, balancing, regularized regression, elastic net, best subset selection

\begin{center}
\end{center}



\baselineskip=20pt\newpage
\setcounter{page}{1}
\renewcommand{\thepage}{\arabic{page}}
\renewcommand{\theequation}{\arabic{section}.\arabic{equation}}


\section{Introduction}

We consider the problem of estimating the causal effect of an intervention in a panel data setting, where we observe the outcome of interest for a number of treated units (possibly only a single one), and a number of control units, for a number of periods  prior to the receipt of the treatment, and for a number of periods after the receipt of the treatment. 
Two aspects of the problem make this different from standard analyses of causal  effects using matching approaches (see \citet{imbenswooldridge} for a recent review).
First, the key variables on which we try to match  treated and control units are pre-treatment outcomes rather than qualitatively different characteristics.
Second, in social science applications the setting is often one where  the number of control units, as well as the number of pre-treatment periods for which we observe outcomes are modest, and of similar magnitude. In fact, a substantial number of applications has only a single, or very few, control units so that estimators motivated by consistency arguments that rely on a large number of control units can have poor properties.

Many of the modern methods researchers have used in this setting can be divided into  three broad groups. First, difference-in-differences (DID) methods (e.g., \citet{ashenfelter1985using, cardmariel, meyer1995workers, abadie2005semiparametric, Bertrand2004did}) where the difference in average pre-treatment outcomes between treated and control units is subtracted from the difference in average post-treatment outcomes between treated and control units, with generalizations to multiple factor structures in \citet{xu2015generalized} and \citet{gobillon2013regional}. Second, matching methods where, for each  treated unit, one or more matches are found among the controls,  based on both pre-treatment outcomes and other covariates (e.g., \citet{abadie2006, diamond2013genetic, rubin2006matched, heckman1997matching, heckman1998matching}). Third, synthetic control (SC)  methods
(\citet{abadie2003, abadie2010, abadie2014, hainmueller}), where for each treated unit a synthetic control is constructed as a weighted average of control units such that the weighted averages matches pre-treatment outcomes and covariates for the treated units.

In this paper we develop new methods for this setting. We make two specific contributions. First, we develop a   framework  that nests many of the existing approaches. In this framework we characterize the estimated counterfactual outcome for the treated unit as a linear combination of outcomes for the control units. This framework allows researchers to contrast the critical assumptions underlying the previously proposed methods. Substantive differences between applications, and differences in the data configurations may make some methods more appropriate in some cases than in others.
For example,  a key difference between DID on the one hand, and matching and  SC approaches on the other hand, is that the DID approach allows for a non-zero intercept in this linear representation, corresponding to a permanent additive difference between the treatment and control units. Such an additive difference is often found to be important in empirical work. 
Second, DID methods restrict the weights on the control units to be equal, whereas matching and SC methods allow variation in weights to capture the notion that some control units make better matches for the treated unit than others. Furthermore, many of the current methods, including DID, matching (other than kernel matching with higher order kernels) and SC impose non-negativity of the weights. All restrict the weights to sum to one.

In a second contribution
 we  propose a  new estimator that relaxes a number of the restrictions specific to, or common among, the DID, matching,  and SC methods.
 Generalizing DID methods we allow the weights to vary. Generalizing SC and matching methods we allow for permanent additive effects. Generalizing DID, SC and nearest neighbor matching methods we allow the weights to be non-negative and do not restrict the weights to sum to one.
   Our proposed method can accommodate cases with many or few controls, and with many or few pre-treatment periods. In the latter case there is a need for regularization or shrinkage, although standard $L_1$ (lasso) type shrinkage towards zero is not necessarily appropriate in general, and in particular if we wish to impose a restriction on the sum of the weights. Specifically we recommend an approximate balancing method with an elastic net penalty term for the weights.

We illustrate the proposed methods using three data sets used previously in this literature.

\section{Notation}\label{notation}
We consider a panel data setting in which there are $N+1$ cross-sectional units observed in time periods $t=1,\dots,T$. 
There is a subset, possibly containing only a single unit, of treated units. For ease of exposition we focus on the case with a single treated unit, unit $0$. From period $T_0$ onwards, for $1<T_0\leq T$, this unit receives the treatment of interest.
Using the potential outcome or Rubin Causal Model set up (\citet{rubin1974estimating, holland1986statistics, imbens2015causal}), there are for the treated unit, in each of the periods $t=T_0$ through $t=T$  a pair of potential outcomes $Y_{0,t}(0)$ and $Y_{0,t}(1)$, corresponding to the outcome given the  control and active treatment respectively. 
The causal effects for this unit for each time period are $\tau_{0,t}=Y_{0,t}(1)-Y_{0,t}(0)$, for $t=T_0+1,\ldots,T$.

Units $i=1,\dots,N$ are control units which do not receive the treatment in any of the time periods. For these units there is a control outcome $Y_{i,t}(0)$, but not necessarily a treated potential outcome. In many examples conceptualizing a treated outcome for the control units can be difficult -- for exampe, in one of the canonical synthetic control applications to the German re-unification, it is difficult to imagine the treated (re-unified with East Germany) state for  countries other than Germany -- and we do not need to do so. 
The treatment received is denoted by $W_{i,t}$, satisfying:
\[W_{i,t}=
\left\{
\begin{array}{ll}
1\hskip1cm & {\rm if}\  i=0, \ {\rm and} \ t\in\{T_0+1,\ldots,T\},\\
0 & {\rm otherwise.}\end{array}\right.\]
We are interested in the treatment effects for the unit who receives the treatment, during the period this unit receives the treatment, that is, $\tau_{0,t}$, for $t=T_0+1,\ldots,T$.

The researcher observes, for unit $i$ in period $t$, the treatment $W_{i,t}$ and the realized outcome, $Y^\obs_{i,t}$: 
\[Y^\obs_{i,t}=Y_{i,t}(W_{i,t})=
\left\{
\begin{array}{ll}
Y_{i,t}(0)\hskip1cm & {\rm if}\  W_{i,t}=0,\\
Y_{i,t}(1) & {\rm if}\ W_{i,t}=1.\end{array}\right.\] 
The researcher may also observe $M$ time-invariant individual-level characteristics $X_{i,1},\dots,X_{i,M}$ for all units.

In the following discussion  we denote by $X_i$ the $M\times 1$ column vector $(X_{i,1},\dots,X_{i,M})^\top$, for  $i=0,\dots,N$. This vector may also include some of the lagged outcomes, $Y^\obs_{i,t}$, in periods $t\leq T_0$. We denote by $\bxn$ the $N\times M$ matrix with the $(i,m)^{\rm th}$ entry equal to $X_{i,m}$, for $i=1,\ldots,N$ and $m=1,\ldots,M$, excluding the treated unit,  and denote by $\bxe$ a $M$-row vector with the $m^{\rm th}$ entry equal to $X_{0,m}$, and finally $\bx=(\bxn,\bxe)$. Similarly, for the outcome, $Y^\obs_i$ denotes the $T\times 1$ vector $(Y^\obs_{i,T},\dots,Y^\obs_{i,1})^\top$.  In addition $\byzn$ denotes the $ N\times T_0$ matrix with the $(i,t)$th entry equal to $Y^\obs_{i,T_0-t+1}$, again excluding the treated unit,
 $\byze$ denotes a $ T_0$-vector with the $t$-th entry equal to $Y^\obs_{0,t}$, and similarly for  $\byen$  and  $\byee$ for the post-treatment period.
The elements of the three matrices $ \byen$, $\byze$, and $\byzn$   consist of observations of the control outcome $Y_{i,t}(0)$, and $\byee$ consists of observations of the treated outcome $Y_{i,t}(1)$.
Combining these matrices we have
\[ \by=\left(
\begin{array}{cc}
\byee & \byen\\
\byze & \byzn
\end{array}
\right)=\left(
\begin{array}{cc}
\btyee & \bnyen\\
\bnyze & \bnyzn
\end{array}
\right),\hskip1cm {\rm and}\ \ 
\bx=\left(
\begin{array}{cc}
\bxe & \bxn
\end{array}
\right).\]

The causal effect of interest depends on the pair of matrices $\btyee$ and $\bnyee$. The former is observed, but the latter is not. Putting aside for the moment the presence of  covariates,
the question is how to use the three different sets of control outcomes,  $ \bnyen$, $\bnyze$, and $\bnyzn$, and specifically  how to model their joint relation with the unobserved  $\bnyee$ in order to impute the latter:
\[ \bz(0)=\left(
\begin{array}{cc}
? & \bnyen\\
\bnyze & \bnyzn
\end{array}
\right).\] 

  One approach is to model the relationship between   $\bnyze$ and $\bnyzn$, and assume that this relation is the same as that between 
$\bnyee$ and $\bnyen$. This is where the current setting is fundamentally different from that where the pre-treatment variables are fixed characteristics rather than pre-treatment outcomes: modelling the relation between covariates for the treated unit and the control units would not necessarily  translate into a prediction for the post-treatment outcome for the treated unit given post-treatment outcomes for the control units.
An alternative approach 
 is to model the relationship between   $\bnyen$ and $\bnyzn$, and assume that this relation is the same as that between 
$\bnyee$ and $\bnyze$. 

To put the problem, as well as the  estimators that we discuss in this paper in context, it is useful to bear in mind the relative magnitude of the different dimensions, the number of control units $N$ and the number of pre-treatment periods $T_0$. 
Part of the motivation to pursue one particular identification strategy, rather than a different one,  may be the relative magnitude of the different components of $\by$, and the corresponding ability, or lack thereof, to precisely estimate their relationship.
Put differently, depending on these relative magnitudes there may be a need for regularization in the estimation strategy and a more compelling case to impose restrictions that are typically viewed as approximations at best.

Sometimes we have few pre-treatment time periods but relatively many control units, $N>>T_0$, e.g.,  
\[ \bz(0)=\left(
\begin{array}{c|cccccc}
?  & Y_{1,3}(0) & Y_{2,3}(0) & Y_{3,3}(0) & Y_{4,3}(0) & \hdots & Y_{N,3}(0)  \\ \hline
Y_{0,2}(0)  & Y_{1,2}(0) & Y_{2,2}(0) & Y_{3,2}(0) & Y_{4,2}(0) & \hdots & Y_{N,2}(0)  \\
Y_{0,1}(0)  & Y_{1,1}(0) & Y_{2,1}(0) & Y_{3,1}(0) & Y_{4,1}(0) & \hdots & Y_{N,1}(0)  \\
\end{array}
\right).\]
In this case it is difficult to estimate precisely the dependence structure between 
$\bnyze$ and $\bnyzn$, relative to the 
 dependence between $\bnyen$ and $\bnyzn$.
In this case simple matching methods are attractive. Matching methods suggest looking for one or more controls that are each similar to the treated unit. With $T_0$ small, there are few dimensions where the units need to be similar, and with $N$ large, we have  a large reservoir of controls to draw from.

Other times the researcher may have relatively many pre-treatment periods but  few control units, 
$T_0>>N$,
e.g.,
\[ \bz(0)=\left(
\begin{array}{c|cc}
?  & Y_{1,T_0+1}(0) & Y_{2,T_0+1}(0)  \\ \hline
Y_{0,T_0}(0)  & Y_{1,T_0}(0) & Y_{2,T_0}(0)  \\
Y_{0,T_0-1}(0)  & Y_{1,T_0-1}(0) & Y_{2,T_0-1}(0)  \\
\vdots&\vdots&\vdots\\
Y_{0,2}(0)  & Y_{1,2}(0) & Y_{2,2}(0)  \\
Y_{0,1}(0)  & Y_{1,1}(0) & Y_{2,1}(0)  \\
\end{array}
\right).\]
In this case there is little chance of finding a control unit among the small reservoir of controls that is similar to the treated unit in all of the many pre-treatment periods. 
Instead it may be easier to estimate precisely the dependence structure between 
$\bnyze$ and $\bnyzn$, for example using an autoregressive model. 
This may motivate time-series approaches as in \citet{brodersen2015inferring} and  \citet{von2015causal}.

In other cases the magnitudes of the cross-section and time series dimension may be similar, $T_0\approx N$. In that case the choice between strategies may be more difficult, and a regularization strategy  for limiting the number of control units that enter into the estimation of $Y_{0,T_0+1}(0)$ may be crucial:
\[ \bz(0)=\left(
\begin{array}{c|cccc}
?  & Y_{1,T_0+1}(0) & Y_{2,T_0+1}(0) & \hdots & Y_{N,T_0+1}(0) \\ \hline
Y_{0,T_0}(0)  & Y_{1,T_0}(0) & Y_{2,T_0}(0) & \hdots & Y_{N,T_0}(0) \\
\vdots&\vdots&\vdots&\ddots&\vdots\\
Y_{0,2}(0)  & Y_{1,2}(0) & Y_{2,2}(0)  & \hdots & Y_{N,2}(0)\\
Y_{0,1}(0)  & Y_{1,1}(0) & Y_{2,1}(0)  & \hdots & Y_{N,1}(0)\\
\end{array}
\right).\]
Finding a cross-section unit that is similar to the treated unit in all of the pre-treatment periods is again likely to be difficult in this case. It may be easier to find a linear combination of controls that is similar to the treated unit in all pre-treatment outcomes.
As a result combinations of  cross-section approaches as in the traditional DID literature (e.g. \citet{ashenfelter1985using, cardmariel, cardkrueger1, meyer1995workers, angristkruegerstrategies, Bertrand2004did, 
imai2016should, angristpischke, atheyimbenscic}), and time-series approach as in \citet{brodersen2015inferring} and  \citet{von2015causal} may be useful, but some type of regularization may be called for. 

\section{Four Leading Applications}\label{fourapplications}

To frame the discussion of the estimators discussed in Sections \ref{generalmethod} and \ref{meth},  let us briefly review four influential  applications from the DID and SC literatures. In particular 
we wish to give a sense of the relative magnitudes of the control sample size $N$ and the number of pre-intervention periods $T_0$, to make the point that primarily relying on consistency under large $N$ or large $T_0$ may not be credible.

\subsection{The Mariel Boatlift Study}

One of the classic applications of DID methods is  the Mariel Boatlift study by
\citet{cardmariel}. 
Card studies the effect of the influx of low-skilled labor into the Miami labor market on wages using data on  labor markets in other metropolitan areas for comparison.
Recently this study has been revisited using synthetic control methdos in   \citet{peri2015}. 
The \citet{peri2015} study uses a single treated unit, 
$N=44$ potential control units, $T_0=7$ pre-treatment periods and $T_1=6$ post-treatment periods.

\subsection{The New-Jersey Pennsylvania Minimum Wage Study}

In the seminal
\citet{cardkrueger1} study, the focus is on the causal effect of a change in the minimum wage in New Jersey. Card and Krueger use data from fast food restaurants in New Jersey and Pennsylvania.
They  use information on $N=78$ control (Pennsylvania) units, 321 treated (new Jersey) units, one pre-treatment period, $T_0=1$, and one post treatment period, $T_1=1$.

\subsection{The California Smoking Legislation Study}

In the seminal study on SC methods, \citet{abadie2010} focus on estimating the effect of  anti-smoking legislation in California. It uses smoking per capita as the outcome and uses  a single treated unit (California) and $N=29$ states without such anti-smoking measures as the set of potential controls. 
\citet{abadie2010} use information on $T_0=17$ pre-program years and data on $T_1=13$ post-program years.

\subsection{The German Re-Unification Study}

 In another classic SC application, 
\citet{abadie2014} study the effect on per capita Gross Domestic Product in West-Germany of the re-unification with East Germany. They use a single treated unit (West-Germany), $N=16$ countries as potential controls and use $T_0=30$ years of data prior to re-unification and $T_1=14$ years of data post re-unification.

\section{A Class of Estimators}\label{generalmethod}

In this section we focus on the setting without covariates. The goals is to impute the 
unobserved control outcomes for the treated unit,  $\bnyee$,  on the basis of three sets of control outcomes,  the pre-treatment period outcomes for both treated and control units,  and the post-treatment period outcomes for the control units,
$ \bnyen$, $\bnyze$, and $\bnyzn$. We then use these imputed values to estimate
the causal effect $\tau_{0,t}$ of the receipt of the treatment on the outcome for unit $0$ in time periods $t=T_0+1,\dots,T_0+T_1$.

\subsection{A Common Structure}

Let us focus on the causal effect for unit $0$ and  for period $T$ for the moment, $\tau_{0,T}=Y_{0,T}(1)-Y_{0,T}(0)$. Because this unit  receives the active treatment during these periods, it follows that
$Y^\obs_{0,T}=Y_{0,T}(1)$, and therefore
the causal effect is equal to $\tau_{0,T}=Y^\obs_{0,T}-Y_{0,T}(0)$, with only $Y_{0,T}(0)$ unobserved. The first observation we make is that many of the estimators in the literature share the following linear structure for the imputation of the unobserved $Y_{0,T}(0)$:
\begin{align}\label{general}
\hat{Y}_{0,T}(0) = {\mu} + \sum_{i=1}^{N} \omega_i \cdot Y^\obs_{i,T}.
\end{align}
In other words, the imputed control outcome for the treated unit is a linear combination of the control units, with intercept $\mu$ and weight $\omega_i$ for control unit $i$.\footnote{One exception is the Changes-In-Changes (CIC) method, a  nonlinear generalization of the linear DID method, developed in \citet{atheyimbenscic}. Another exception is \citet{brodersen2015inferring} which develops a Bayesian method that allows for time-varying coefficients in the regression.}
 The various  methods differ in the way the parameters in this linear combination, the intercept $\mu$ and the weights $\omega$,  are chosen as a function of the outcomes $\byen$, 
$\byze$, and $\byzn$ (but typically not involving $\byee$). 
One obvious way to choose the parameters $\mu$ and $\omega$, given the characterization in (\ref{general}),  is to estimate them by least squares:
\begin{equation}\label{ols1} (\hat\mu^\ols,\hat\omega^\ols)=
\arg\min_{\mu,\omega}
\sum_{s=1}^{T_0} \left( Y^\obs_{0,s} -\mu-\sum_{i=1}^{N} 
\omega_i\cdot Y^\obs_{0,s}\right)^2.\end{equation}
This regression involves $T_0$ observations and $N+1$ predictors (the $N$ potential control units and an intercept). 
This approach may be attractive in settings where the number of pre-treatment outcomes $T_0$ is large relative to the number of control units $N$, but would be less so in cases where they are of similar magntitude. As illustrated by the examples in Section \ref{fourapplications}, in practice $T_0$ and $N$ are often of a magnitude that simply estimating this regression by least squares is not likely to have good properties. 
In its basic form it may not even be feasible if the number of control units is larger than the number of pre-treatment periods. Even if the number of pre-treatment periods is large enough to make this approach formally feasible, the resulting estimator may suffer from lack of precision. This leads to a need for some regularization  for, or restrictions on, the weights $\omega$.

\subsection{Four Restrictions on the Intercept and Weights}
\label{constraints}

Here 
we focus on the representation (\ref{general}) of $\hat{Y}_{0,T}(0) $ as a linear combination of outcomes for the control units.
We discuss  four constraints on the parameters, both the intercept $\mu$ and the weights $\omega$, that have been considered in the literature. In general none of these restrictions are likely to hold in practice and we propose relaxing all four of them. However, relaxing all of them can create problems with statistical precision, leading to a need for statistical regularization.

The  four constraints we consider are:
\begin{align}
  \mu &= 0,\tag{NO-INTERCEPT}\label{adh.1}\\
  \sum_{i=1}^N \omega_i &= 1,\tag{ADDING-UP}\label{adh.2}\\
 \omega_i&\geq 0,\ i=1,\dots,N,\tag{NON-NEGATIVITY}\label{adh.3}\\
\omega_i & = \overline{\omega},\ i=1,\dots,N.\tag{CONSTANT-WEIGHTS}\label{con.did}
\end{align}

The first three  restrictions are imposed by \citet{abadie2010, abadie2014}
in  the original synthetic control analyses of the California smoking and the Germany re-unification applications. 
Single nearest neighbor matching (\citet{abadie2006}) also imposes the first three restrictions, and in addition allows $\omega_i$ to differ from zero for only a single control unit.
More general  matching methods such as kernel matching  and bias-corrected matching (e.g., \citet{heckman1997matching, heckman1998matching, abadie2011bias} allow for multiple non-zero  and  negative weights. 
The standard DID methodology imposes the last three restrictions.

The first restriction, \ref{adh.1} rules out the possibility that the outcome for the treated unit is systematically larger, by a constant amount, than the other units. Note that allowing for such a systematic additive difference between the treatment unit and the control units is a critical feature of the standard DID strategy, which assumes that the trends in the control outcomes in the different groups  are parallel, but which allows for permanently different levels for  the different units.

The second restriction, \ref{adh.2}, requires that the weights sum up to one. It is common to  matching, DID and SC methods. Like the no-intercept restriction, however, this restriction is implausible if the unit of interest is an outlier relative to the other units. For example, 
in the California smoking example, if the outcome was total number of cigarettes smoked in the state, this would be implausible because California is by far the largest US state in terms of population. Using per capita smoking rates as the outcome instead of total number of cigaretts, as  \citet{abadie2010} in fact do, addresses part of this problem, but it may still not be sufficient to make this restriction plausible.
Taking the first two restrictions together, however, makes it difficult to obtain good predictions for  extreme units, that is, units with systematically the largest or smallest values for the outcome.

The third restriction, \ref{adh.3},  requires the weights to be nonnegative.  
This is a key restriction in the ADH estimator, playing  a dual role in their approach. It helps regularize the estimation of the weights in cases with relatively many control units by ensuring in many cases that there is  a unique solution. It also helps control the precision of the resulting imputation by limiting the sum of the squared weights which enters into the variance. Finally, it often ensures that the weights are non-zero only for a small subset of the control units, making the weights easier to interpret.
The restriction is also substantively interesting. In many cases it is plausible, and verifiable, that the raw correlations between the pre-treatment outcomes for each pair of units are positive. However, this does not mean that the partial correlations are all non-negative, and allowing for negative weights may well improve the out-of-sample prediction.

To illustrate why one might wish to use negative weights, consider a setting where the units are states, with one treated state and two control states, and a key characteristic is the share of young and old people in the state. If the 
share of young people in the treated state is 2/3, and the shares of the young people in the two control states are 1/2 and 1/3, then it may well be that using weights 2 (for the control state with share of the young equal to 1/2) and -1 (for the control state with share of the young equal to 1/3) leads to better results because it would make the synthetic control state have a share of the young equal to 2/3, identical to that in the treated state.

A second reason to allow for negative weights is the role they play in bias-reduction. In nearest neighbor matching (\citet{abadie2006}) the bias goes to zero slowly in settings with many covariates that are to be matched on. Allowing the weights to be negative as in bias-corrected matching estimators  (\citet{abadie2011bias}), allows one to improve this rate. Note that even simple least squares estimators for treatment effects with covariates implicitly allow for negative weights on the control units.

In cases with the number of potential control units $N$ larger than the number of pre-treatment outcomes $T_0$, and especially when $N$ is much larger than $T_0$, the combination of the first  three restrictions need not lead to a unique set of values for $\mu$ and $\omega$. In such cases there might be multiple  values that satisfy these constraints. We therefore need to find a way of further regularizing the choice of weights, by restricting the set, or by ranking the parameter values within the set of values that satisfy the constraints.
There are a number of ways of doing so that have been proposed in the literature. 
Matching methods look for a small set of control units, or even just a single one, that are similar to the treated unit. Part of the motivation for this is that it may be more credible to approximate a treated unit by pretreatment values $(2,3)$ with a single control with pretreatment values $(2,3)$ rather than as a combination of two controls, one with pretreatment values $(1,1)$ and one with pretreatment values $(3,5)$. The reason is that the approximation with two control units relies more on the linearity being an accurate approximation.
A second approach is to  use the 
 fourth restriction, \ref{con.did}, which strengthens the nonnegativity condition by making the assumption that all control units are equally valid. This assumption,  standard in DID analyses, suggests combining the control units by  setting all weights equal. In combination with restriction \ref{adh.2} this implies that the weights are all equal to $1/N$. Relaxing this restriction is a critical feature of the SC approach.

It is important to stress that the allowing the intercept to differ from zero is conceptually different here, than it is in standard matching settings, because in the latter approach the matching or balancing is on covariates that are qualitatively different from lagged outcomes. Consider the California smoking example where the outcome is number of cigarettes per capita. Suppose we have two covariates, beer consumption and cigarette prices. It does not make sense to look for a linear combination of other states with an intercept such  that the linear combination of the other states matches California both in terms of beer consumption and in terms of cigarette prices. Even if there was such a linear combination, so that, both for beer consumption and cigarette prices, California is equal to 3+0.8$\times$UT+0.5$\times$TX, the results  would not be scale-invariant: changing prices from dollars to cents would imply that the linear approximation would not longer be valid even after changing the coefficients. 
With the covariates qualitatively different the linear model would only make sense if the weights sum to one, and if there is no intercept.
When all the covariates are lagged outcomes, however, allowing for a non-zero intercept, and allowing the sum of the weights to deviate from one, does not violate scale invariance because
all covariates would change by the same factor.

\subsection{The Objective Function}

There may be many pairs of $(\mu,\omega)$ that satisfy the set of restrictions imposed, and in fact we may not wish to impose all, or even any, of the restrictions. Within the set of $(\mu,\omega)$ that satisfy the restrictions imposed we consider rankings of the pairs of values that take the form of preferences over $\omega$. In general we prefer values such that the synthetic control unit is similar to the  treated unit in terms of lagged outcomes. In addition, we prefer values such that the dispersion of the weights is small. We may also prefer to have few control units with non-zero weights, although this is a more controversial objective to formally justify. Here we discuss a specific objective function, although other approaches are possible.

The first component of the objective function focuses on balance between the treated unit and the control units. Specifically it focuses on the difference between the pre-treatment outcomes for the treated unit and the linear combination of the pre-treatment outcomes for the control units:
\begin{align}
\left\|\byze-\mu-\omega^\top \byzn\right\|_2^2=\left(\byze-\mu-\omega^\top \byzn\right)^\top
\left(\byze-\mu-\omega^\top \byzn\right)
.\tag{BALANCE}\label{obj.1}\end{align}
If $T_0$ is sufficiently large relative to $N$, we may be able to find values for $(\mu,\omega)$ that uniquely minimize this objective functions. However, in many cases this will not be possible, a finding  also noted in \citet{abadie2016}. When there are multiple weights that provide an exact solution to 
\begin{equation}
\byze=\mu+\omega^\top \byzn,\label{con.balance}
\end{equation}
we need to use an objective function that directly compares different values of the weights, in other words, we need to regularize the estimator for $\omega$.

The  second component of the objective function  does so by focusing on the values of the weights themselves. 
There are two components to the objective function, which implicitly captures a preference for  small number of non-zero weights, as well as explicitly a preference for smaller weights:
\begin{align*}
&\|\omega\|_1= \sum_{i=1}^{N}|\omega_i|,\hskip1cm {\rm and}\ \ 
\|\omega\|_2^2= \sum_{i=1}^{N}\omega_i^2
.
\end{align*}
 We can capture both by using an elastic-net type penalty (\citet{hastie2009elements, hastie2015statistical}) that combines these Lasso  and ridge terms:
\begin{align}
\lambda\cdot\left(\frac{1-\alpha}{2} \|\omega\|_2^2+
\alpha\|\omega\|_1\right)
.\tag{PENALTY FUNCTION}\label{obj.2}
\end{align}
In \citet{brodersen2015inferring} the authors take a Bayesian approach, and use a spike and slab prior distribution (\citet{george1997approaches}) to deal with the potentially large number of parameters.

Alternatively one might want to add a penalty term of the form $\|\omega\|_0=\sum_{i=1}^N {\bf 1}_{\omega_i\neq 0}$, directly penalizing the number of non-zero weights. Such a penalty, typically in the form of directly restricting 
$\sum_{i=1}^N {\bf 1}_{\omega_i\neq 0}$, is implicit in the matching literature.

\subsection{The Proposed Method}
\label{proposedmethod}

Our main recommendation is to  estimate the intercept and weights as
\[ \left(\hat\mu^\elastic(\lambda,\alpha),\hat\omega^\elastic(\lambda,\alpha)\right)
=\arg\min_{\mu,\omega}Q\left(\mu,\omega\left|\byze,\byzn;\lambda,\alpha\right.\right)
,\]
where
\begin{align*}
Q\left(\mu,\omega\left|\byze,\byzn;\lambda,\alpha\right.\right)=\left\|\byze-\mu-\omega^\top \byzn\right\|_2^2+
\lambda\cdot \left(\frac{1-\alpha}{2}\|\omega\|_2^2+
\alpha\|\omega\|_1\right),
\tag{OBJECTIVE FUNCTION}\label{obj.3}\end{align*}
without imposing any of the four restrictions  (\ref{adh.1})-(\ref{con.did}). 
The superscript ``en'' here stands for elastic net, referring to the form of the penalty term.
The price for relaxing all four of these restrictions is that we need to impose some regularization on the estimators through the choice of the parameters of the penalty term, $\lambda$ and $\alpha$. 
There are three issues that requires slight modifications to  standard approaches to  regularization here. First, we do not want to scale the covariates $\byzn$, because that would change the interpretation of the weights in cases. Without normalization the restriction that the weights sum up to one is an important substantive one to consider. With the normalization of the covariates this adding up restriction would no longer be substantively meaningful.
Second, the  weights are likely to sum up to a number close to one, so that shrinking towards zero needs to be done with care.
Third, if one actually imposes the exact adding up restriction on the $\omega_i$, as well as the non-negativity constraint, lasso-style $L_1$ regularization does not work because the penalty term would equal  $\sum_{i=1}^N |\omega_i|=\sum_{i=1}^N \omega_i=1$ for all values of the weights considered.

Given these issues
we propose a particular cross-validation procedure, without normalizing the covariates.  
Consider 
 the elastic net procedure with no restrictions on $\mu$ and $\omega$.
We treat each control unit in turn as the pseudo-treated unit, to determine the optimal value for the tuning parameters.
When we use unit $j$ as the pseudo-treated unit, given tuning parameters $\alpha$ and $\lambda$, 
this leads to a set of weights  $\hat{\omega}^\elastic_i(j;\alpha,\lambda)$ and an intercept $\hat{\mu}^\elastic(j;\alpha,\lambda)$:
\[\left(\hat{\mu}^\elastic(j;\alpha,\lambda),\hat{\omega}^\elastic(j;\alpha,\lambda)\right)
\]
\[\hskip2cm 
=\arg\min_{\mu,\omega}
\sum_{t=1}^{T_0} \left(Y^\obs_{j,t}-\mu-\sum_{i\neq 0,j} \omega_i\cdot Y^\obs_{i,t}\right)^2+
\lambda\cdot \left(\frac{1-\alpha}{2}\|\omega\|_2^2+
\alpha\|\omega\|_1\right)\]
Given these weights we predict the outcome for unit $j$ in period $T$ as
\[ \hat Y_{j,T}(0)=\hat{\mu}^\elastic(j;\alpha,\lambda)+\sum_{i\neq j} \hat\omega^\elastic_i(j;\alpha,\lambda)\cdot Y^\obs_{i,T}.\]

 The performance of the model is then evaluated by computing the mean squared error, for period $T$,  averaged over all control units
\[
   CV^\elastic(\alpha,\lambda)= \frac{1}{N}\sum_{j=1}^{N}\left(Y^\obs_{j,T}
-\hat{\mu}^\elastic(j;\alpha,\lambda)-\sum_{i\neq 0,j} \hat\omega^\elastic_i(j;\alpha,\lambda)\cdot Y^\obs_{i,T}\right)^2.
\]
We choose the value of the tuning parameter that minimizes the cross-validation error,
\begin{align*}
   \left(\alpha_{\rm opt}^{\elastic},\lambda_{\rm opt}^{\elastic}\right) = \argmin_{\alpha,\lambda} &\ \Bigl\{CV^\elastic(\alpha,\lambda)\Bigr\}.
\end{align*}
We consider a finite set of values for $\alpha\in\{0.1,0.1,\ldots,0.9\}$, and all possible positive values for $\lambda$, $\lambda\in(0,\infty)$.
Given ($\lambda_\opt^\elastic,\alpha_\opt^\elastic$), define the final estimates as
\[ \left(\hat\mu^\elastic,\hat\omega^\elastic\right)
=\arg\min_{\mu,\alpha}Q\left(\mu,\omega\left|\byze,\byzn;\lambda_\opt^\elastic,\alpha_\opt^\elastic\right.\right)
.\]

Although we focus here primarily on the estimator without any of the four restrictions, in some applications one may wish to impose  some of those restrictions on substantive grounds or to improve precision. Here they are not required, however, for regularization purposes.  

\section{Four Alternative Methods}\label{meth}

Here we discuss four alternative methods  for choosing $\mu$ and $\omega$ to put our proposed method in perspective. A number  of these have been previously proposed. The current set up allows for a comparison in a common setting.
The first three of these methods impose subsets of the restrictions (\ref{adh.1})-(\ref{con.did}) that we do not impose. As a result of that they do not have the need for the regularization on the weights that we employ.

\subsection{Difference-in-Differences}\label{did}

The original DID method 
(e.g., \citet{ashenfelter1985using, cardmariel, cardkrueger1, meyer1995workers, angristkruegerstrategies, Bertrand2004did, angristpischke, atheyimbenscic})
can be thought of as solving the optimization problem (\ref{model:general}) subject to
 (\ref{adh.2}), (\ref{adh.3}), and (\ref{con.did}). In other words, it solves
\begin{align}\label{model:general}
  \Bigl(\hat{\mu}^\did,\hat{\omega}^\did\Bigr) &= \argmin_{\mu,\omega}
Q\left(\mu,\omega\left|\byze,\byzn;\lambda=0,\alpha\right.\right)\\
& {\rm s.t.}\ \sum_{i=1}^N \omega_i=1,\ \omega_i\geq 0,\ \omega_i=\overline{\omega}.
\end{align}
imposing the restrictions (\ref{adh.2}), (\ref{adh.3}), and (\ref{con.did}). This implies the $\hat\omega^\did$ do not depend on the data, leading to
\begin{align*}
  \omega_i^\did & = \frac{1}{N},\ i=1,\dots,N,
\\
  \hat{\mu}^\did &= \frac{1}{T_0}\sum_{s=1}^{T_0} Y^\obs_{0,s} - \frac{1}{N T_0}\sum_{s=1}^{T_0}\sum_{i=1}^{N} Y_{i,s}^\obs.
\end{align*}
This in turn leads to estimates for  $Y_{0,t}(0)$, for the periods $t\geq T_0+1$, 
equal to
\begin{align}
  \hat Y_{0,t}^\did(0) &=
\hat\mu^\did+ \sum_{i=1}^{N} \hat\omega_i^\did\cdot Y^\obs_{i,t}\\&=
 \Biggl(\frac{1}{T_0}\sum_{s=1}^{T_0} Y^\obs_{0,s} - \frac{1}{N T_0}\sum_{s=1}^{T_0}\sum_{i=1}^{N} Y^\obs_{i,s}
\Biggr) + \frac{1}{N}\sum_{i=1}^{N} Y^\obs_{i,t}.
\end{align}
Let us consider this in the special case with a single pre-treatment period, $T_0=1$.
In that case there is no unique solution for $(\mu,\omega)$ based on (\ref{ols1}), and 
the DID approach addresses this by fixing $\omega$ at $1/N$, and using the pre-treatment period to estimate $\mu$ as 
$\hat\mu^\did=Y^\obs_{0,1}- \frac{1}{N}\sum_{i=1}^{N} Y_{i,1}^\obs$. This leads to
\[\hat\tau^\did=
\left( Y^\obs_{0,2}- \frac{1}{N}\sum_{i=1}^{N} Y_{i,2}^\obs\right)
-\left( Y^\obs_{0,1} - \frac{1}{N}\sum_{i=1}^{N} Y_{i,1}^\obs
\right) .\]
The constant weights restriction takes care of any need to regularize the estimation of the weights $\omega$. With that restriction  there is a unique solution for $\mu$ even in the case with a single pre-treatment period.
\citet{xu2015generalized} considers generalizations that allow for a more complex factor structure.

\subsection{The Abadie-Diamond-Hainmueller Synthetic Control Method}\label{adh}

The original synthetic control method of \citet{abadie2010} 
imposes the restrictions that the intercept is zero, and that weights are non-negative and sum up to one, (constraints (\ref{adh.1}),  (\ref{adh.2}) and (\ref{adh.3})). The weights $\hat{\omega}^\adh$ are chosen to match both the pre-treatment outcomes and a set of fixed characteristics, denoted by the $M$-component vector $X_i$ for unit $i$. We first discuss the original ADH implementation in the general case with covariates and then return to the special case with no covariates.

Given an $M\times M$ positive semi-definite diagonal matrix $V$, define the weights $\hat\omega(V)$ as the solution
\begin{align}\label{synth.2}
  (\hat{\omega}(V),\hat\mu(V)) = \argmin_{\omega,\mu} &\ \Biggl\{\left(\bxe-\mu-\omega^\top\bxn\right)^\top V\left(\bxe-\mu-\omega^\top\bx\right)\Biggr\}\\
  \text{s.t.} &\ 
\sum_{i=1}^N \omega_i = 1\quad\text{and}\quad \omega_i \geq 0,\ i=1,\dots,N, \ \mu=0\nonumber
\end{align}
These weights  minimize the distance between the treated unit  and the weighted combination of the other units in terms of the covariates $X_i$. (Note that in the general ADH approach these covariates $X_i$ may include some or all of the pre-treatment $Y^\obs_{i,t}$).

The diagonal weight matrix $V$ is then chosen to match the lagged outcomes:
\begin{align}\label{synth.1}
  \hat{V} = \argmin_{V={\rm diag}(v_1,\dots,v_M)} &\ \Biggl\{\left(\byze-\hat{\omega}(V)^\top \byzn\right)^\top \left(\byze-\hat{\omega}(V)^\top \byzn\right)\Biggr\}\\
  \text{s.t.} &\ \sum_{m=1}^M v_m = 1\quad\text{and}\quad v_m \geq 0,\ m=1,\dots,M.\nonumber
\end{align}
The ADH weights are then $\hat\omega^\adh=\hat\omega(\hat V)$ (and $\hat\mu^\adh=0$).
In general the researcher has a choice regarding what to put in the vector of pretreatment variables $X_i$. This vector may include some or all of the pretreatment outcomes $Y^\obs_{i,t}$ for $t=1,\ldots,T_0$.

\subsection{Constrained Regression}\label{constrained}

Now consider the special case of the ADH method  where $X_i$ is equal to the full vector of pretreatment outcomes $Y_{i,t}$ for $t=1,\ldots,T_0$, and contains no other variables. In that case the unconstrained weights that minimize (\ref{synth.1}) are the weights that
solve (\ref{synth.2}) with $V$ equal to the $N\times N$ identity matrix. We refer to this special case of the ADH method as the constrained regression. We can characterize it slightly differently  by fitting it into the general framework
(\ref{model:general}):
\begin{align}\label{conreg}
  \hat{\omega}^{\constr} = \argmin_{\mu,\omega} &
Q\left(\mu,\omega\left|\byze,\byzn;\lambda=0,\alpha\right.\right)\\
\text{s.t.} &\ \ \mu=0,\ \ \ \sum_{i=1}^N \omega_i = 1\quad\text{and}\quad \omega_i \geq 0,\ i=1,\dots,N.\nonumber
\end{align}
The original version of the ADH approach, as described in Section \ref{adh}, makes it clear why it imposes the \ref{adh.1} restriction.
As discussed before in Section \ref{constraints}, in an application with qualitatively different covariates,  it makes little sense to allow there to be a difference between the treated unit and the weighted average of the control units that is the same for different covariates.
In the context where the pretreatment variables are all the same variable, however, just measured at different points in time,  allowing those differences to be different from zero but requiring them to be the same can be a meaningful relaxation, the  way it is in standard DID methods.
For the constrained estimator, therefore, there is no particular reason why one would impose the restriction that the intercept is zero, and this restriction can easily be relaxed. Similarly the adding-up restriction can be relaxed without any problems. Note that we do not claim that one should always relax these restrictions, our point is that these are substantive restrictions that should be considered on their merit. 

Relaxing the zero intercept restriction (\ref{adh.1}), but maintaining the adding-up restriction 
(\ref{adh.2}), makes it easier to compare the constrained regression (which is close to the original ADH estimator) and the standard difference-in-difference approach.
The remaining difference is that the DID imposes the restriction that the weights $\omega_i$  are all identical  (restrictions (\ref{adh.2})
 and (\ref{con.did})), implying that the weights are all equal to $1/N$.
Relaxing this restriction, and allowing the weights to vary, is arguably the key innovation of the ADH approach over the standard DID approach.
In the constrained regression version it becomes clear that this improvement can be achieved without any additional restrictions.
Moreover, we can relax the other restrictions, (\ref{adh.2}) and (\ref{adh.3}), as well, if there is a sufficiently large number of pretreatment periods.

In both the original ADH approach and the constrained regression version, there need not be a unique solution for $\omega$. Because of the non-negativity constraint on the $\omega$ the question whether this is an issue in a specific application is not simply a matter of counting the number of pre-treatment periods and the number of controls, but with a sufficiently large number of control units it is likely that there are multiple solutions. This problem is exacerbated by relaxing the zero-intercept restriction, but it also can arise in the presence of that restriction.

\subsection{Best Subset Selection}

An alternative approach is to select the set of best controls. For a fixed number of controls, say  $k$, the optimal weights solve
\begin{align}\label{subset}
  \Bigl(\hat\mu^\subs,\hat{\omega}^\subs\Bigr) =
\argmin_{\mu,\omega} &
Q\left(\mu,\omega\left|\byze,\byzn;\lambda=0,\alpha\right.\right)
,\\
{\rm s.t.} &\ \sum_{i=1}^N {\bf 1}_{\omega_i\neq 0}\leq k.\nonumber
\end{align}
The tuning parameter of the model is the number of weights that are allowed to be different from zero, $k$. 
Because of the small sample sizes, using cross-validation may not be an attractive way to go in practice. Instead we propose using a prior distribution for the number of non-zero weights, using a Poisson distribution with mean and variance equal to $\beta$. In practice we recommend setting $\beta=3$.

Part of the  differences between this best-subset method, the ADH method and the related constrained regression concerns the restrictions \ref{adh.1} and \ref{adh.2}.
Both the restriction that the intercept is zero, and the restriction that the weights sum up to one can be relaxed easily in the constrained regression. A more important difference between the two methods is the fact that the best subset selection does not require the weights to be non-negative. 
A special case is the best single control
which
 uses the pre-treatment data to select a single control with weight equal to one:
\begin{align}\label{single}
  \Bigl(\hat\mu^\single,\hat{\omega}^\single\Bigr) 
= \argmin_{\mu,\omega} &
Q\left(\mu,\omega\left|\byze,\byzn;\lambda=0,\alpha\right.\right)\\
\text{s.t.} &\ \ \mu=0,\ \ \ \sum_{i=1}^N \omega_i = 1\quad\text{and}\quad \omega_i \geq 0,\ i=1,\dots,N
\ \sum_{i=1}^N {\bf 1}_{\omega_i\neq 0} = 1
.\nonumber
\end{align}
This leads to choosing the control unit $j$ that minimizes
\[ j=\arg\min_{i\in\{1,\ldots,N\}}
\left(\byze-Y^\obs_{i,{\rm pre}}\right)^\top \left(\byze-Y^\obs_{i,{\rm pre}}\right).
\]
In many difference-in-differences applications with a single treatment and single control group, researchers informally choose the control group. The best single control approach formalizes that selection process by choosing the single control unit that is the most similar to the treated unit prior to the treatment similar to matching. One might also wish to relax the restriction that the intercept is zero, to gain flexibility.

\subsection{Covariates}
\label{covariates}

So far the discussion has almost exclusively been about the setting where the only pre-treatment variables were the lagged outcomes. With additional pre-treatment variables there are other issues. 
First, we should note that in practic these other pre-treatment variables tend to play a relatively minor role. In terms of predictive power the lagged outcomes tend to be substantially more important, and as a result the decision how to treat these other pre-treatment variables need not be a a very important one.

As raised in the discussion on the role of the intercept, we cannot treat the pre-treatment variables in the same way as the lagged outcomes. Here we suggest one alternative. Prior to choosing the weights and possibly the intercept, we can regress the control outcomes on the pre-treatment variables and calculate the residuals. Then we use the residuals in the approaches discussed in   Section \ref{proposedmethod}.

\section{Inference}
\label{inference}

To  conduct classical inference one needs to be explicit about what is random in the repeated sampling procedure. 
This is often controversial in synthetic control applications. Especially if there is only a single treated unit, it is often the case that this unit is unique in some aspects. We encountered this issue already in the formal definition of the causal effects and the potential outcomes. There it was useful to be careful not to define $Y_{i,t}(1)$ for control units for whom it may be different to imagine the treated state (e.g., the German re-unification example). 
Here similarly we do not want to argue that there was some positive probability that control units could have received the treatment.

Here we discuss two specific methods for doing inference. In the first case the unit that is treated is viewed as exchangeable with the other units in the absence of the treatment, and in the second case the period in which the treated unit first receive the treatment is stochastic. We also discuss a method for combining the two methods. This type of randomization inference is in the spirit of  the way p-values are calculated in \citet{abadie2010, firpo, ando2013hypothesis}, although here we focus on standard errors rather than p-values. See also \citet{hahn2016, pinto} for a discussion in settings with a large number of pre-treatment periods.

In general the different estimators for $\tau$ can be be written as 
\[ \hat\tau=Y^\obs_{0,T}-\hat Y_{0,T}(0).\]
Because the treatment effect is $\tau=Y_{0,T}(1)- Y_{0,T}(0)=
Y^\obs_{0,T}- Y_{0,T}(0)$, the estimation error arises only from the imputation error for $Y_{0,T}(0)$, that is, 
$\hat\tau-\tau=Y_{0,T}(0)- \hat Y_{0,T}(0)$. Hence the
 squared error  is 
\[ \left(\hat\tau-\tau\right)^2=\left(Y_{0,T}(0)- \hat Y_{0,T}(0)\right)^2.\]

For discussing inference it is useful to have a more general notation for the estimators.
First, we use $\bz_{i,s}^{j,t}(0)$, for $i\leq j$ and $s\leq t$ as shorthand for the matrix
where we use units from the $i$-th unit up to the $j$-th unit and time periods from the $s$-th time period up to the $t$-th time period:
\[ \bz_{i,s}^{j,t}=
\left(\begin{array}{ccc}
Y_{i,t}(0) &\hdots & Y_{j,t}(0)\\
\vdots & \ddots & \vdots\\
Y_{i,s}(0) &\hdots & Y_{j,s}(0)\\
\end{array}
\right),\]
and $\bz^{(i),t}_{(i),s}$ as shorthand for the matrix where we leave out unit $i$ from the matrix with all units, $\bz_{0,s}^{N,t}$:
\[ \bz_{(i),s}^{(i),t}=
\left(\begin{array}{cccccc}
Y_{0,t}(0) &\hdots  & Y_{i-1,t}(0) & Y_{i+1,t}(0)&\hdots& Y_{N,t}(0)\\
\vdots & \ddots & \vdots& \vdots & \ddots & \vdots\\
Y_{0,s}(0) &\hdots  & Y_{i-1,s}(0) & Y_{i+1,s}(0)&\hdots& Y_{N,s}(0)\\
\end{array}
\right).\]

Now suppose that we wish to predict $Y_{i,t}(0)$. There are three set of data that will be used to do so. First, outcome values for unit $i$ in periods $1$ through $t-1$, contained in $\bz^{i,t-1}_{i,1}$. Second, the period $t$ outcomes for other units, $\bz^{(i),t}_{(i),t}$, and third, the earlier outcomes for other units,  $\bz^{(i),t-1}_{(i),1}$.
The estimators for the missing $Y_{0,T}(0)$ discussed so far can be written as functions of these three matrices:
\[ \hat Y_{0,T}(0)=g\left(\bz^{0,T-1}_{0,1}, \bz^{(0),T}_{(0),T}, \bz^{(0),T-1}_{(0),1}\right).\]

\subsection{Random Assignment of the Unit}

In the first approach to doing inference we view the treated unit as exchangeable with the control units in the absence of the treatment. We estimate the variance by analyzing the data as if we were estimating $Y_{i,T}(0)$ for one of the control units. 
Had we estimated $Y_{i,T}(0)$ for  unit $i$ using the same estimator, we would have estimated it 
as
\[\hat Y_{i,T}(0)= g\left(\bz^{i,T-1}_{i,1}, \bz^{(0,i),T}_{(0,i),T}, \bz^{(0,i),T-1}_{(0,i),1}\right).\]
We actually observe $Y_{i,T}(0)=Y_{i,T}^\obs$, and so we can calculate the squared error $(Y_{i,T}(0)-\hat Y_{i,T}(0))^2$, which, if the treated unit was randomly selected, an unbiased estimator for the expected squared error, and thus, ignoring the bias, for the variance. We can do this for all control units, leading to
\begin{equation}\hat\mmv_{c}=\frac{1}{N}\sum_{i=1}^N \left(
Y_{i,T}(0)-g\left(\bz^{i,T-1}_{i,1}, \bz^{(0,i),T}_{(0,i),T}, \bz^{(0,i),T-1}_{(0,i),1}\right)
\right)^2.\end{equation}
This is our preferred estimator for the variance and the one we use in the applications.

We can weaken the random selection of the treated unit assumption by making it conditional on some set of covariates or lagged outcomes. For that to be meaningful we would have to have a substantial number of control units.

\subsection{Random Selection of the Treatment Period}

An alternative is to view the period in which the treated unit was initially treated  as randomly selected. This leads to
\[\hat\mmv_{t}=\frac{1}{s}\sum_{t=T_0-s+1}^{T_0} \left(
Y_{i,t}(0)-g\left(\bz^{0,t-1}_{i,1}, \bz^{(0),t}_{(0),t}, \bz^{(0),t-1}_{(0),1}\right)
\right)^2.\]

\subsection{Combining the Methods}

Finally, we can combine the two approaches, leading to
\[\hat\mmv_{ct}=\frac{1}{N\cdot s}\sum_{i=1}^N \sum_{t=T_0-s+1}^{T_0}\left(
Y_{i,t}(0)-g\left(\bz^{i,t-1}_{i,1}, \bz^{(0,i),t}_{(0,i),t}, \bz^{(0,i),t-1}_{(0,i),1}\right)
\right)^2.\]

\section{Three Applications}
\label{applications}

We use data from  three of the seminal studies in this literature, the California smoking example from \citet{abadie2010},  the West Germany re-unification example, from \citet{abadie2014}, and the Mariel boatlift (\citet{cardmariel, peri2015}).
In all three cases we report five estimates. First the original ADH estimator. Second, the constrained regression modification of the ADH estimator. Third we report the standard DID estimator. The last two estimators are new estimators. We report both the elastic net and the best subset estimator, in both cases without imposing the assumptions \ref{adh.1}, \ref{adh.2}, \ref{adh.3}, or \ref{con.did}.
The goal is to compare the relative performance of the five estimators, and to assess the importance of relaxing the restrictions.

\subsection{The California Smoking Application}

\begin{figure*}[!htbp]
  \centering
  \caption{\textbf{Tobaco Control Program in California}}\label{fg:ca_1}
  \begin{subfigure}{1.0\textwidth}
    \centering
    \includegraphics[width=0.8\linewidth]{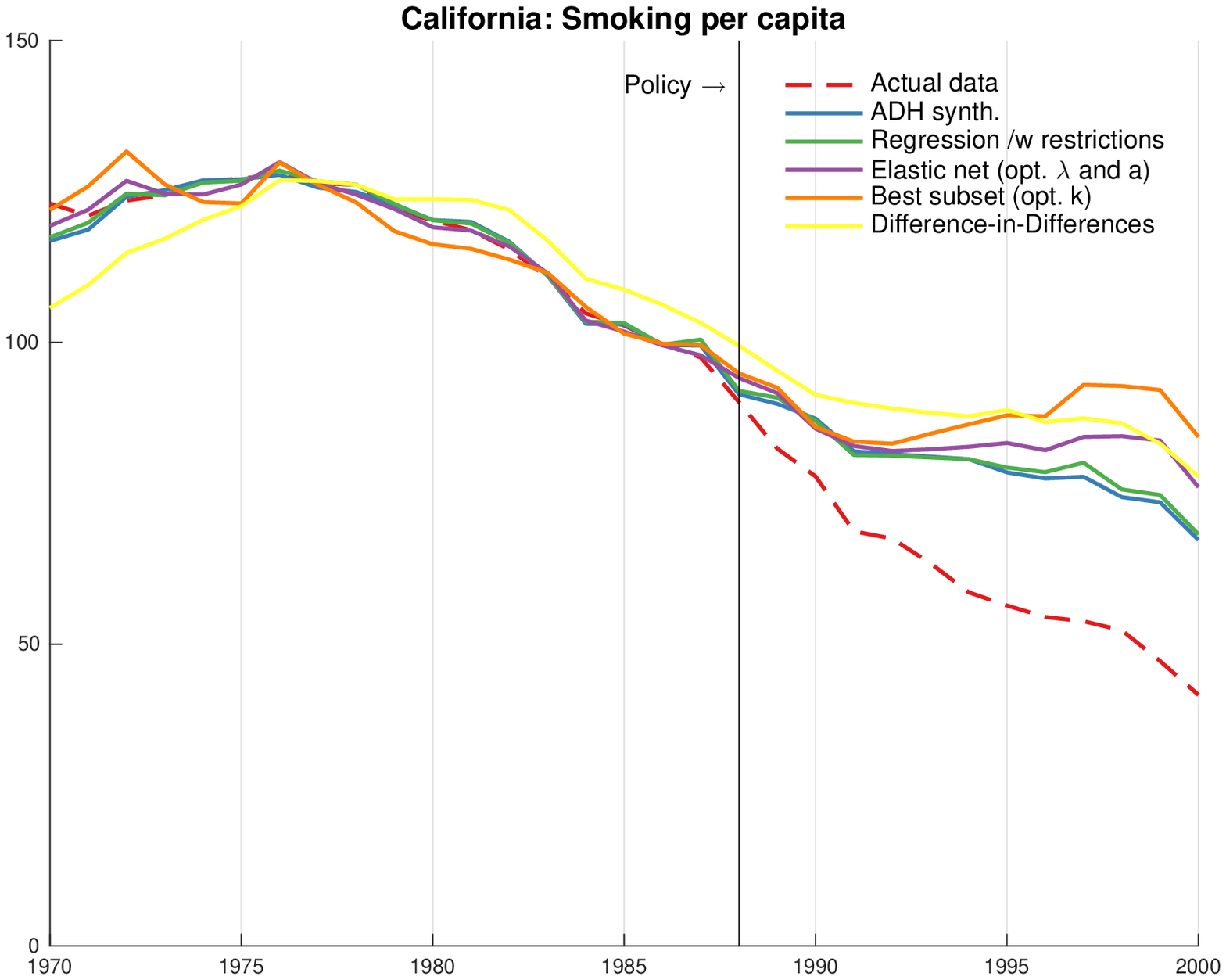}
  \end{subfigure}\\
  \medskip
  \begin{subfigure}{1.0\textwidth}
    \centering
    \includegraphics[width=0.8\linewidth]{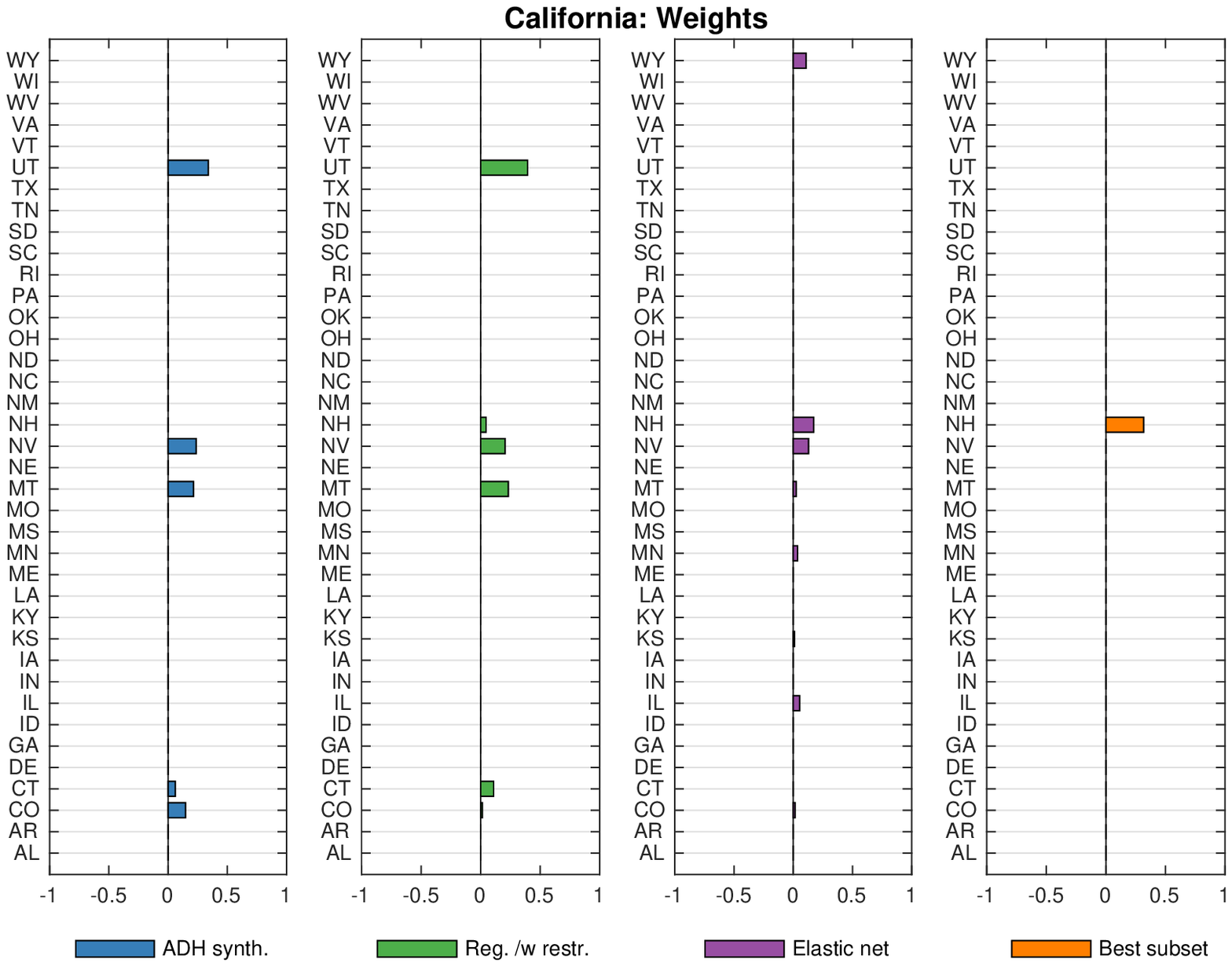}
  \end{subfigure}
\end{figure*}

 \citet{abadie2010} analyze the effect of anti-smoking legislation in California, enacted in January 1989. 
We re-analyze their data using the methods discussed in this paper.
The outcome of interest is the per capita smoking rate. We use data from 1970 to 2000. In Figure \ref{fg:ca_1} we present the actual per capita smoking rate in California, as well as the per capita smoking rate for a synthetic control version of California, constructed using the five estimators discussed in this paper. These five estimators include the original ADH estimator, the constrained estimator with the same restrictions, $\mu=0$, $\sum_{i=1}^N \omega_i=1$ and $\omega_i\geq 0$, the best subset estimator, and DID estimator, and the elastic net estimator. For the best subset estimator the optimal number of controls, based on cross-validation, is 1. For the elastic net estimator the tuning parameters, choosen by cross-validation, are $\alpha=0.1$ and $\lambda=45.5$, leading to 8 states with non-zero weights, all of them positive.
\begin{table}[!htbp]
\begin{center}
\caption{\textbf{California: Parameters}}\label{tb:ca_par}
  \begin{tabular}{l|c|c|c|c}
    \hline\hline
    Model & $\sum_i w_i$ & $\mu$ & $\hat\tau_{{\rm Cal},1995}$ & s.e.\\
    \hline
    Original synth. & $1$ & $0$ & $-22.1$ & $16.1$ \\
    Constrained reg. & $1$ & $0$ & $-22.9$ & $12.8$ \\
    Elastic net & $0.55$ & $18.5$ & $-26.9$ & $16.8$ \\
    Best subset & $0.32$ & $37.6$ & $-31.5$ & $20.3$ \\
    Diff-in-diff & $1$ & $-14.4$ & $-32.4$ & $18.9$ \\ 
    \hline
  \end{tabular}
\end{center}
\end{table}

\clearpage

\begin{figure*}[!htbp]
  \centering
  \begin{subfigure}{1.0\textwidth}
    \centering
    \includegraphics[width=0.8\linewidth]{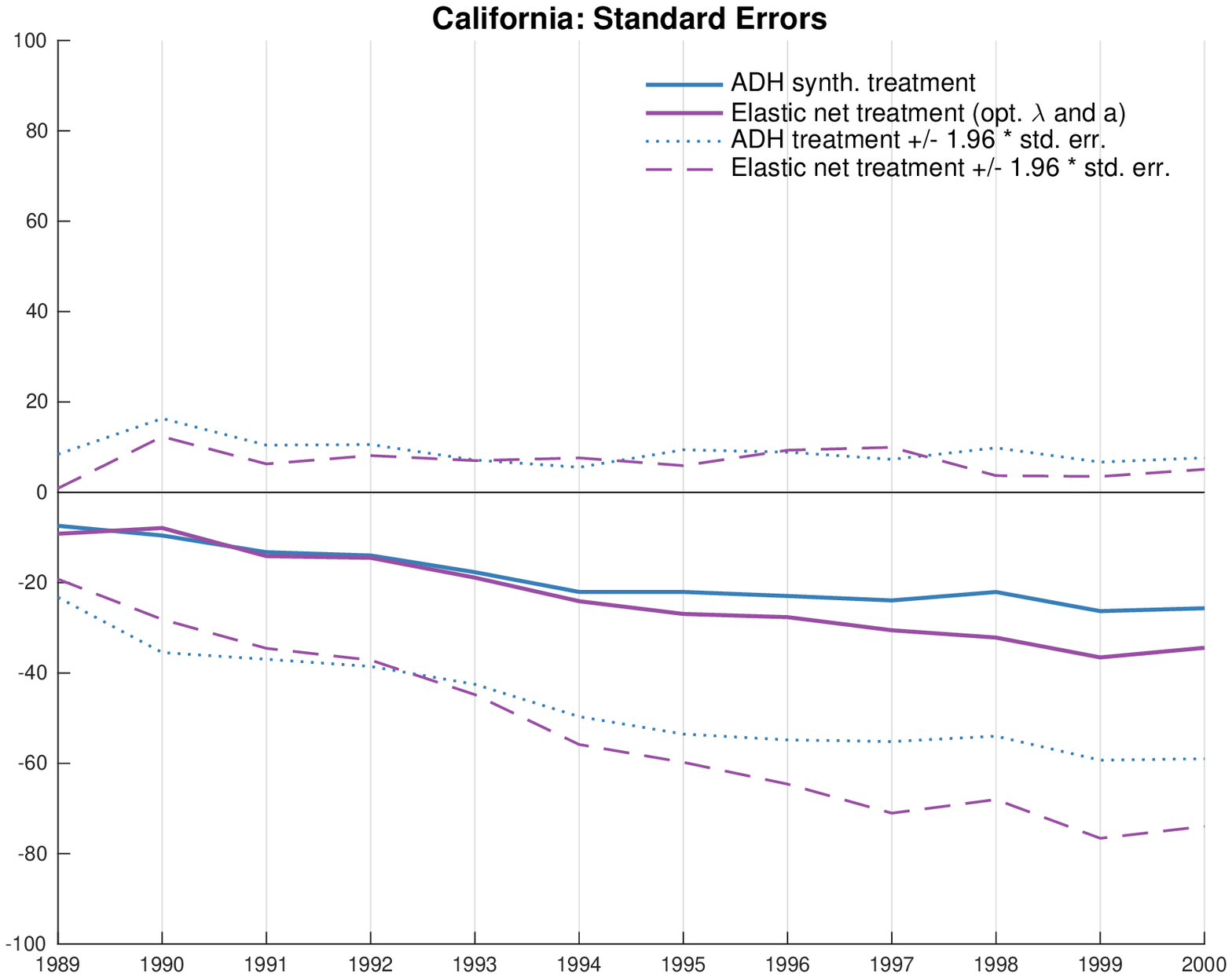}
  \end{subfigure}\\
  \medskip
  \begin{subfigure}{1.0\textwidth}
    \centering
    \includegraphics[width=0.8\linewidth]{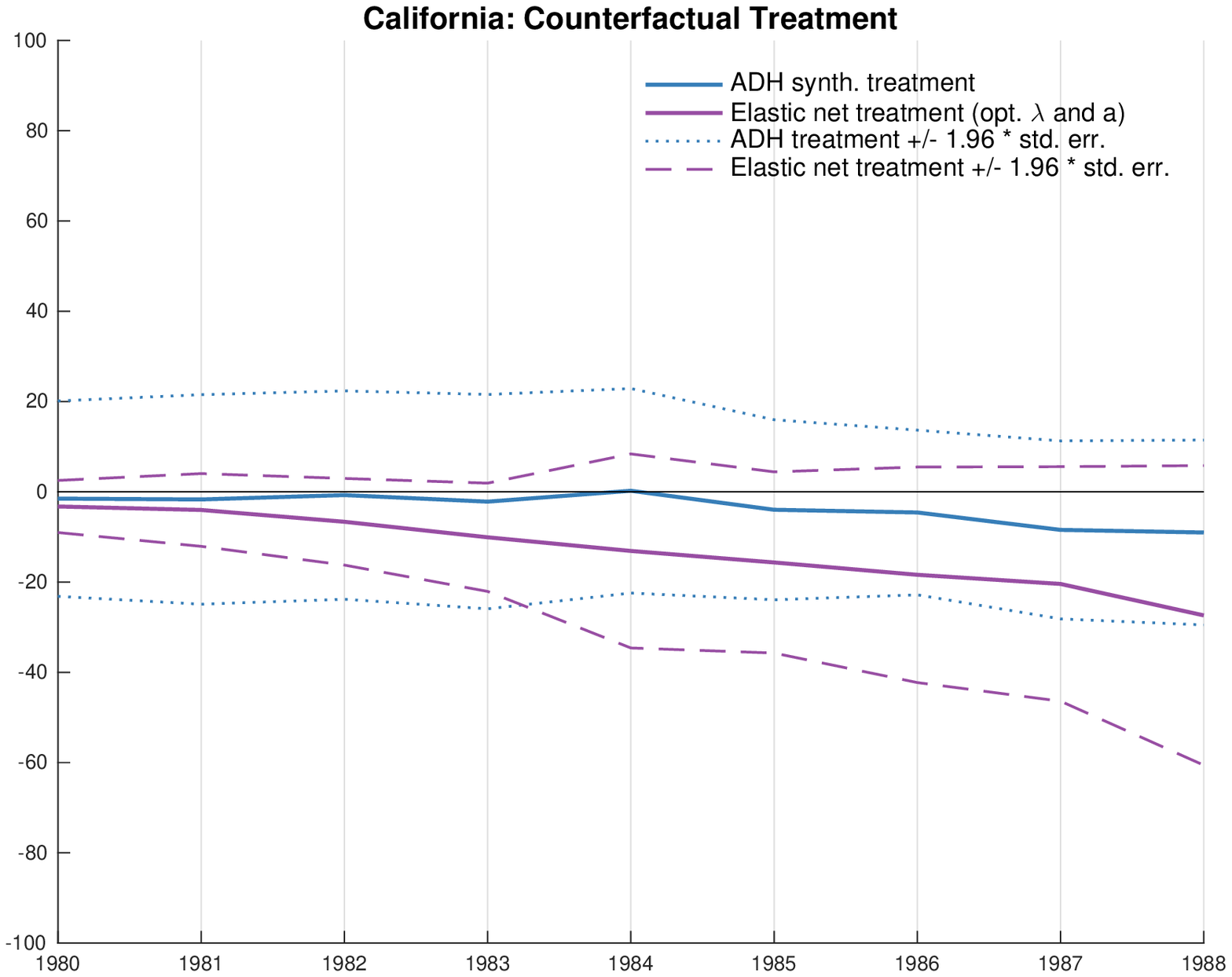}
  \end{subfigure}
\end{figure*}
\clearpage

\subsection{The West Germany Re-Unification Application}

In the second application we revisit the analysis by \citet{abadie2014} of the effect of the German re-unification on West Germany's economy. The outcome is per capita GDP, with data from 1960 to 2004. We compare the same set of five estimators. Here the best subset estimator selects 5 control countries. For the 
elastic net estimator the tuning parameters, choosen by cross-validation, are $\alpha=0.4$ and $\lambda=52.8$, leading to 13 countries with non-zero weights, 2 of them negative.
\begin{table}[!htbp]
\begin{center}
\caption{\textbf{West Germany: Parameters}}\label{tb:wde_par}
  \begin{tabular}{l|c|c|c|c}
    \hline\hline
    Model & $\sum_i w_i$ & $\mu$ & $\hat\tau_{{\rm Ger},1995}$ & s.e.\\
    \hline
    Original synth. & $1$ & $0$ & $-1217$ & $1882.3$ \\
    Constrained reg. & $1$ & $0$ & $-790$ & $1157.6$ \\
    Elastic net & $0.93$ & $213.5$ & $-882$ & $1147.2$ \\
    Best subset & $1.01$ & $168.5$ & $-1019$ & $1364.1$ \\
    Diff-in-diff & $1$ & $1074.1$ & $990$ & $2874.8$ \\ 
\hline  
  \end{tabular}
\end{center}
\end{table}

\begin{figure*}[!htbp]
  \centering
  \caption{\textbf{Reunification of Germany}}\label{fg:wde_1}
  \begin{subfigure}{1.0\textwidth}
    \centering
    \includegraphics[width=0.8\linewidth]{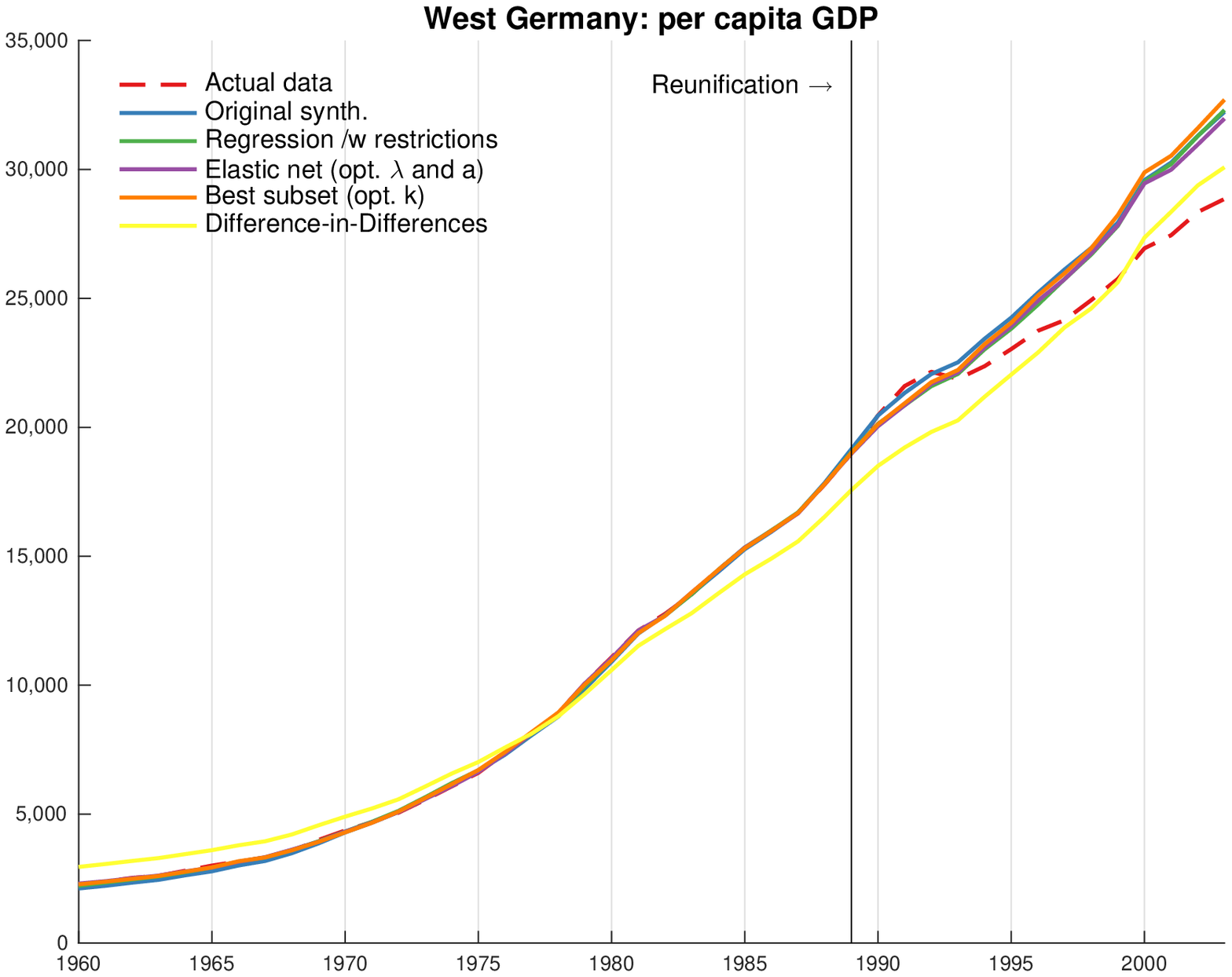}
  \end{subfigure}\\
  \medskip
  \begin{subfigure}{1.0\textwidth}
    \centering
    \includegraphics[width=0.8\linewidth]{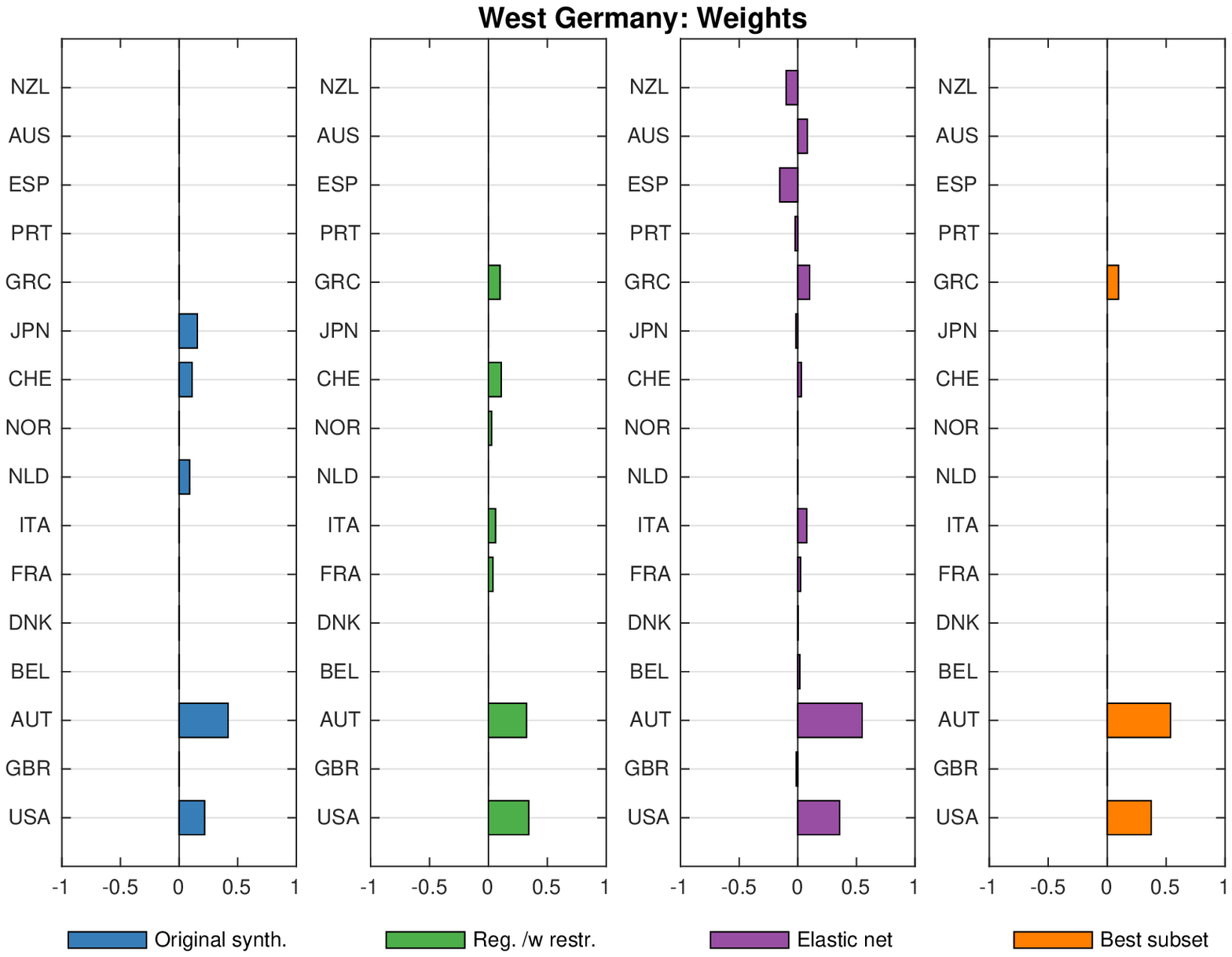}
  \end{subfigure}
\end{figure*}
\clearpage

\begin{figure*}[!htbp]
  \centering
  \begin{subfigure}{1.0\textwidth}
    \centering
    \includegraphics[width=0.8\linewidth]{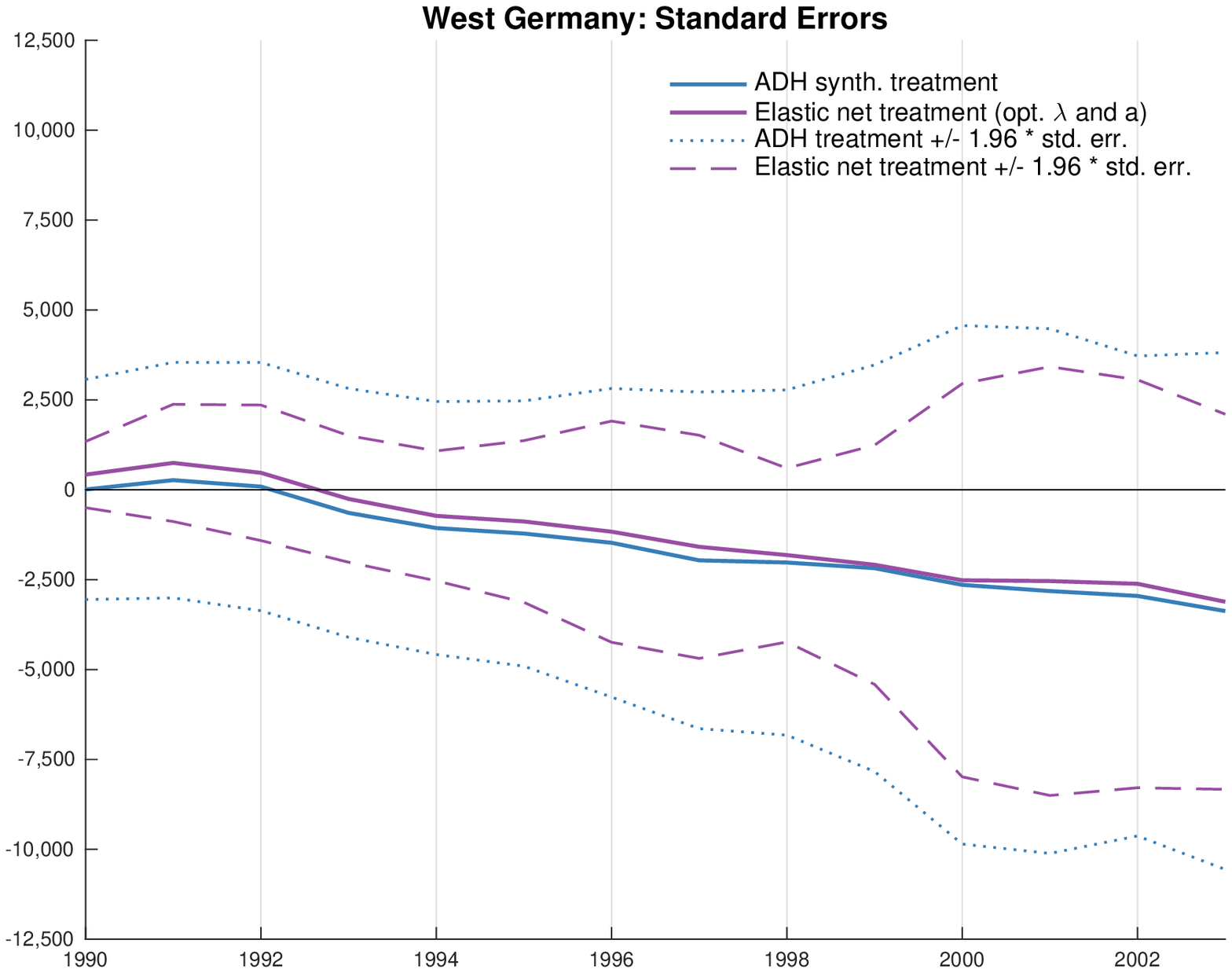}
  \end{subfigure}\\
  \medskip
  \begin{subfigure}{1.0\textwidth}
    \centering
    \includegraphics[width=0.8\linewidth]{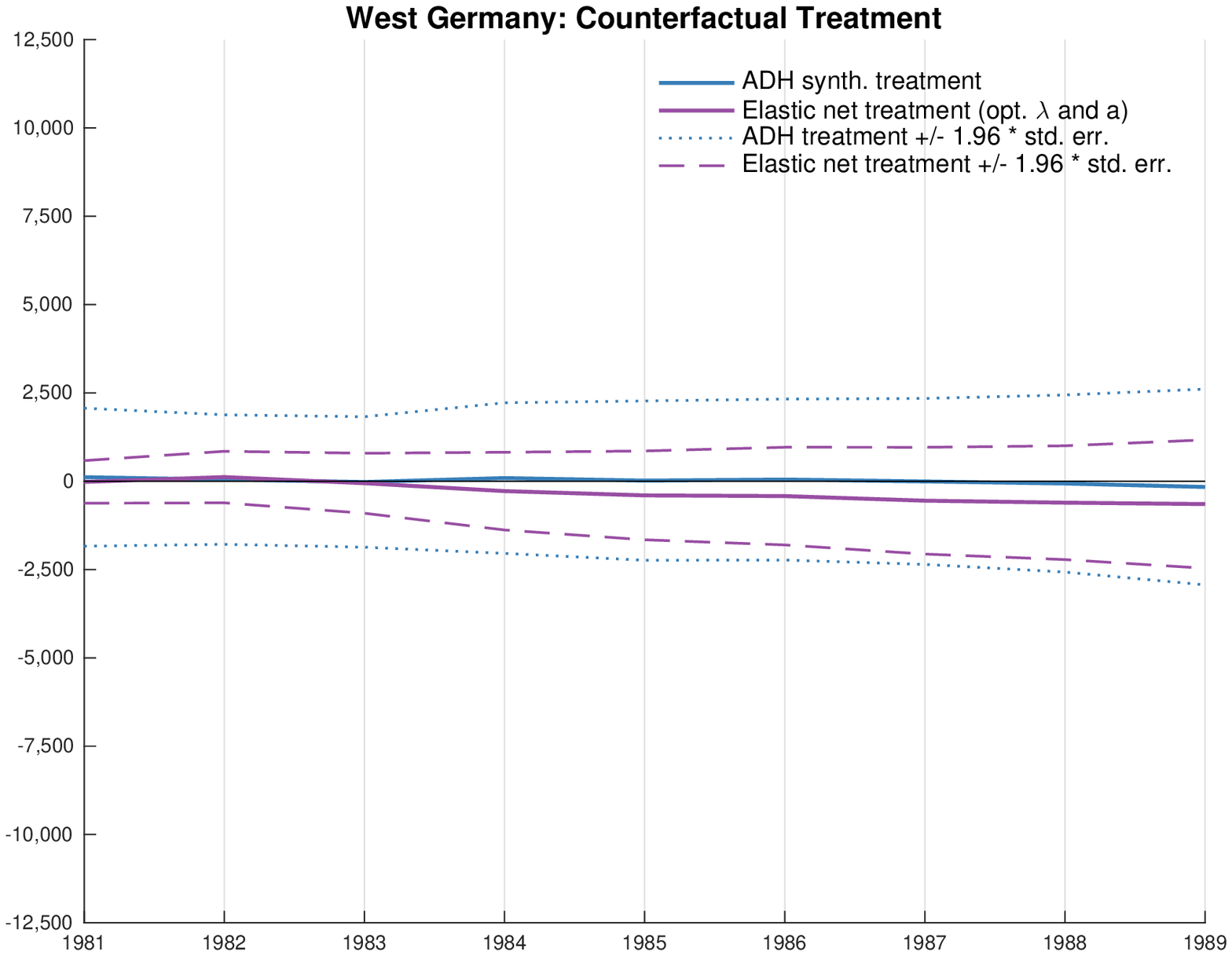}
  \end{subfigure}
\end{figure*}
\clearpage

\subsection{The Mariel Boatlift Application}
In the final application we analyze the effect of Mariel Boatlift on the logarithm of weekly wages using the data from \citet{peri2015}.\footnote{For the counterfactual exercise we drop the average of the logarithm weekly wages and the 1978 logarithm weekly wages from the set of covariates used in the original synthetic control procedure.} Table \ref{tb:boat_par} and Figure \ref{fg:wde_1} report the results obtained for the subpopulation from 16 to 61 years old. For the best subset estimator the cross-validation optimal number of controls is 1. Elastic net selects 22 control units (the optimal tuning parameters are $\alpha = 0.2$ and $\lambda = 0.001$).

\begin{table}[!htbp]
\begin{center}
\caption{\textbf{Mariel Boatlift: Parameters}}\label{tb:boat_par}
  \begin{tabular}{l|c|c|c|c}
    \hline\hline
    Model & $\sum_i w_i$ & $\mu$ & $\hat\tau_{{\rm Miami},1985}$ & s.e.\\
    \hline
    Original synth. & $1$ & $0$ & $0.21$ & $0.19$ \\
    Constrained reg. & $1$ & $0$ & $0.36$ & $0.20$ \\
    Elastic net & $0.37$ & $3.13$ & $0.11$ & $0.27$ \\
    Best subset & $0.69$ & $1.53$ & $0.39$ & $0.44$ \\
    Diff-in-diff & $1$ & $-0.04$ & $0.19$ & $0.17$ \\ 
    \hline
  \end{tabular}
\end{center}
\end{table}

\begin{figure*}[!htbp]
  \centering
  \caption{\textbf{Mariel Boatlift}}\label{fg:wde_1}
  \begin{subfigure}{1.0\textwidth}
    \centering
    \includegraphics[width=0.8\linewidth]{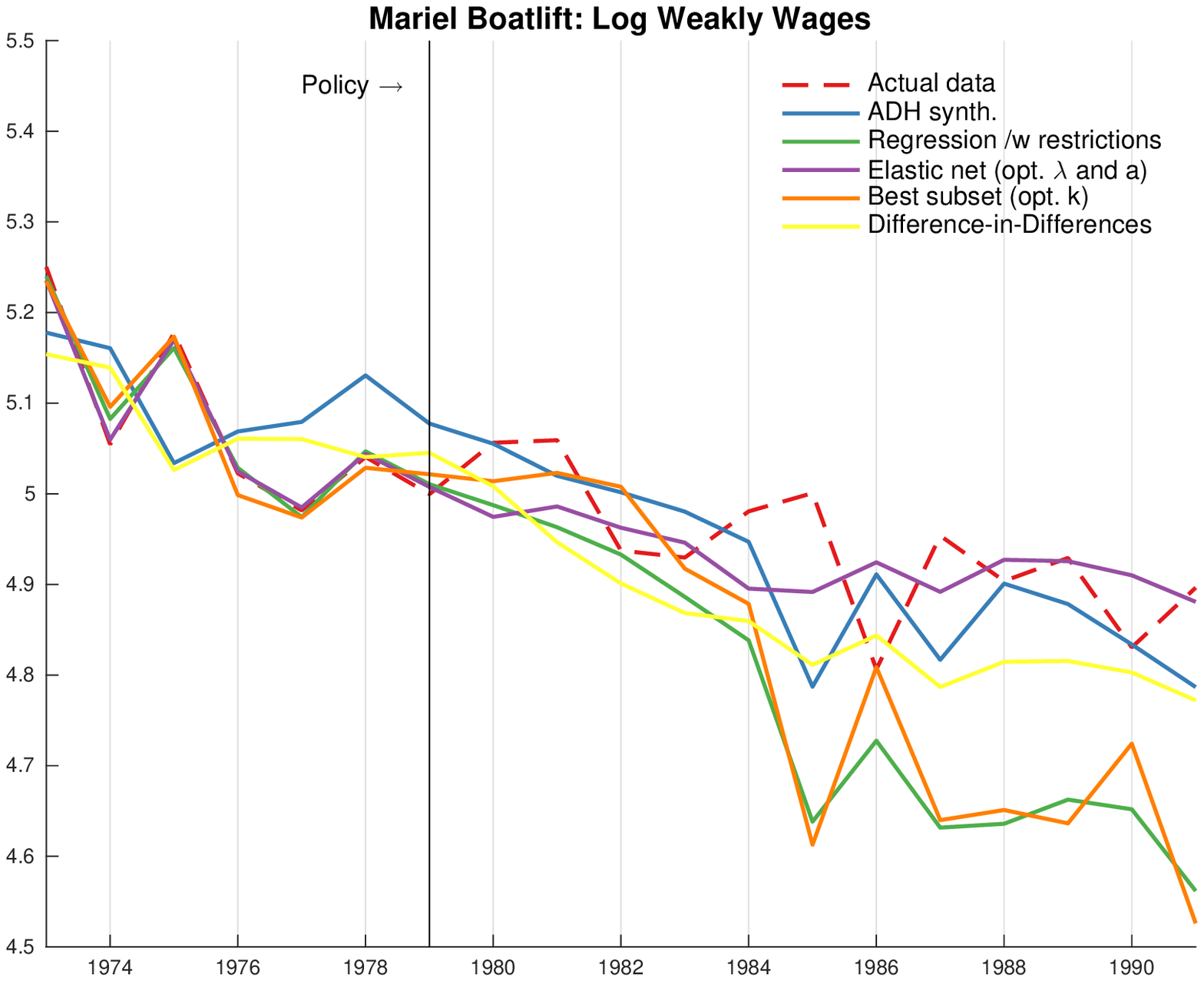}
  \end{subfigure}\\
  \medskip
  \begin{subfigure}{1.0\textwidth}
    \centering
    \includegraphics[width=0.8\linewidth]{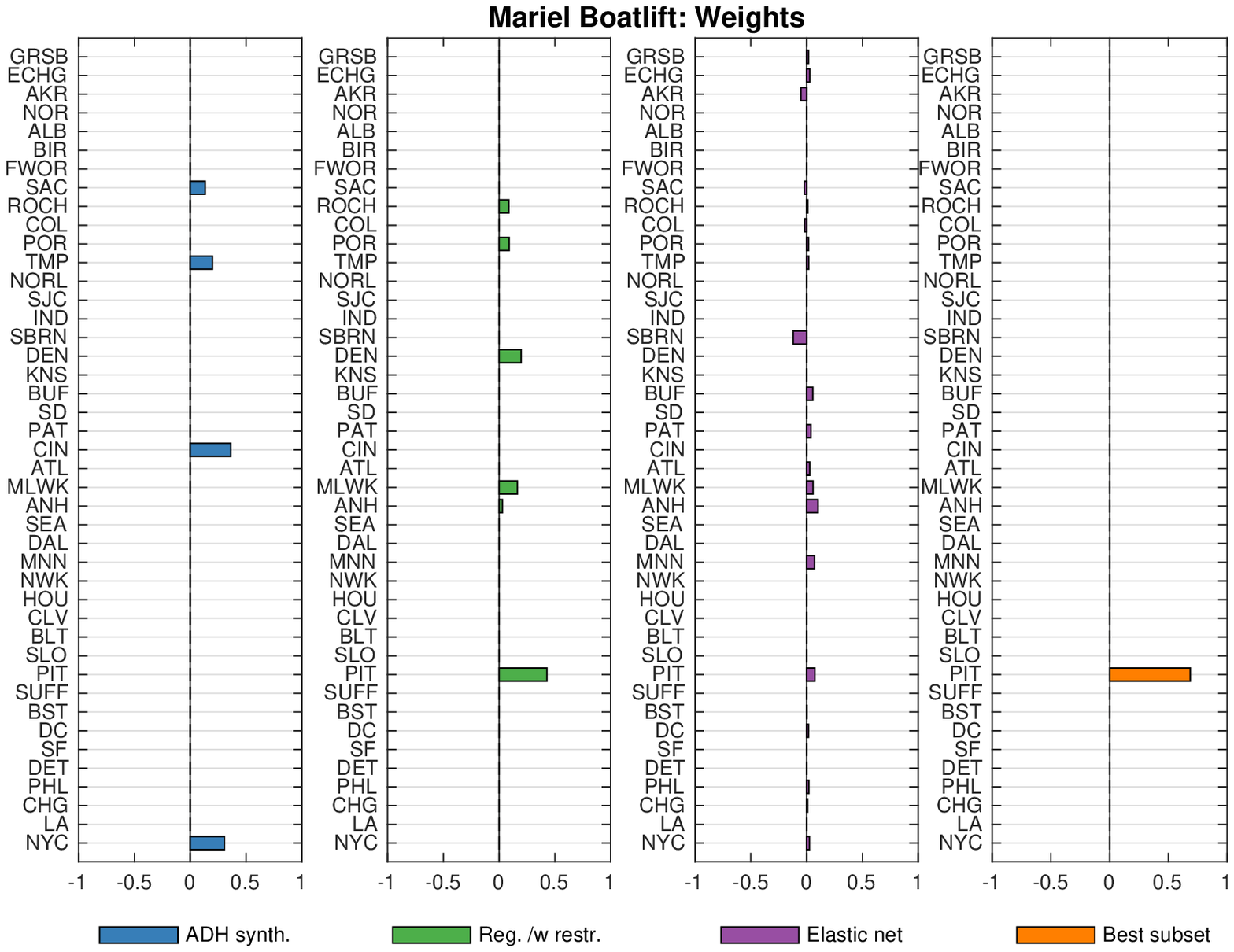}
  \end{subfigure}
\end{figure*}
\clearpage

\begin{figure*}[!htbp]
  \centering
  \begin{subfigure}{1.0\textwidth}
    \centering
    \includegraphics[width=0.8\linewidth]{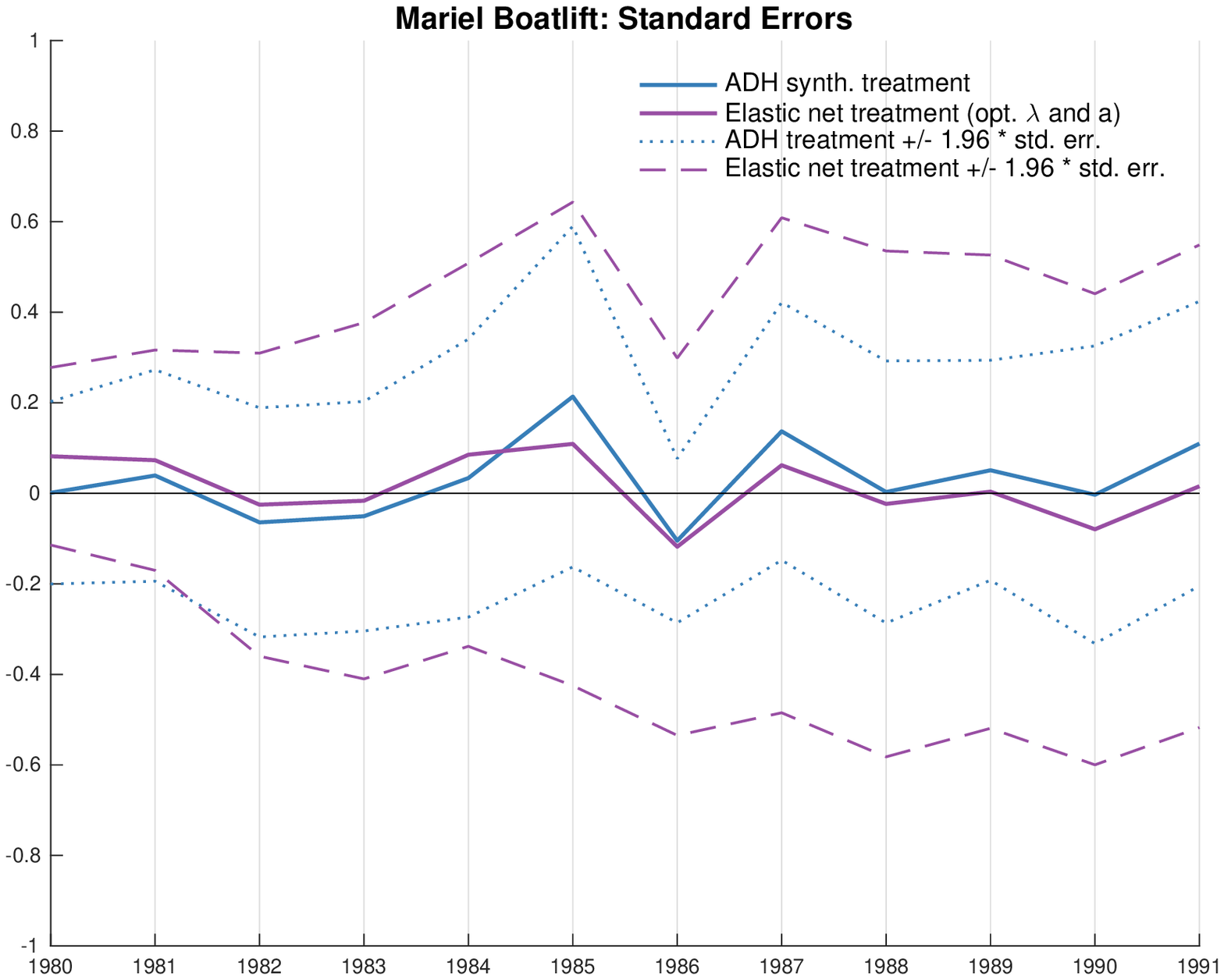}
  \end{subfigure}\\
  \medskip
  \begin{subfigure}{1.0\textwidth}
    \centering
    \includegraphics[width=0.8\linewidth]{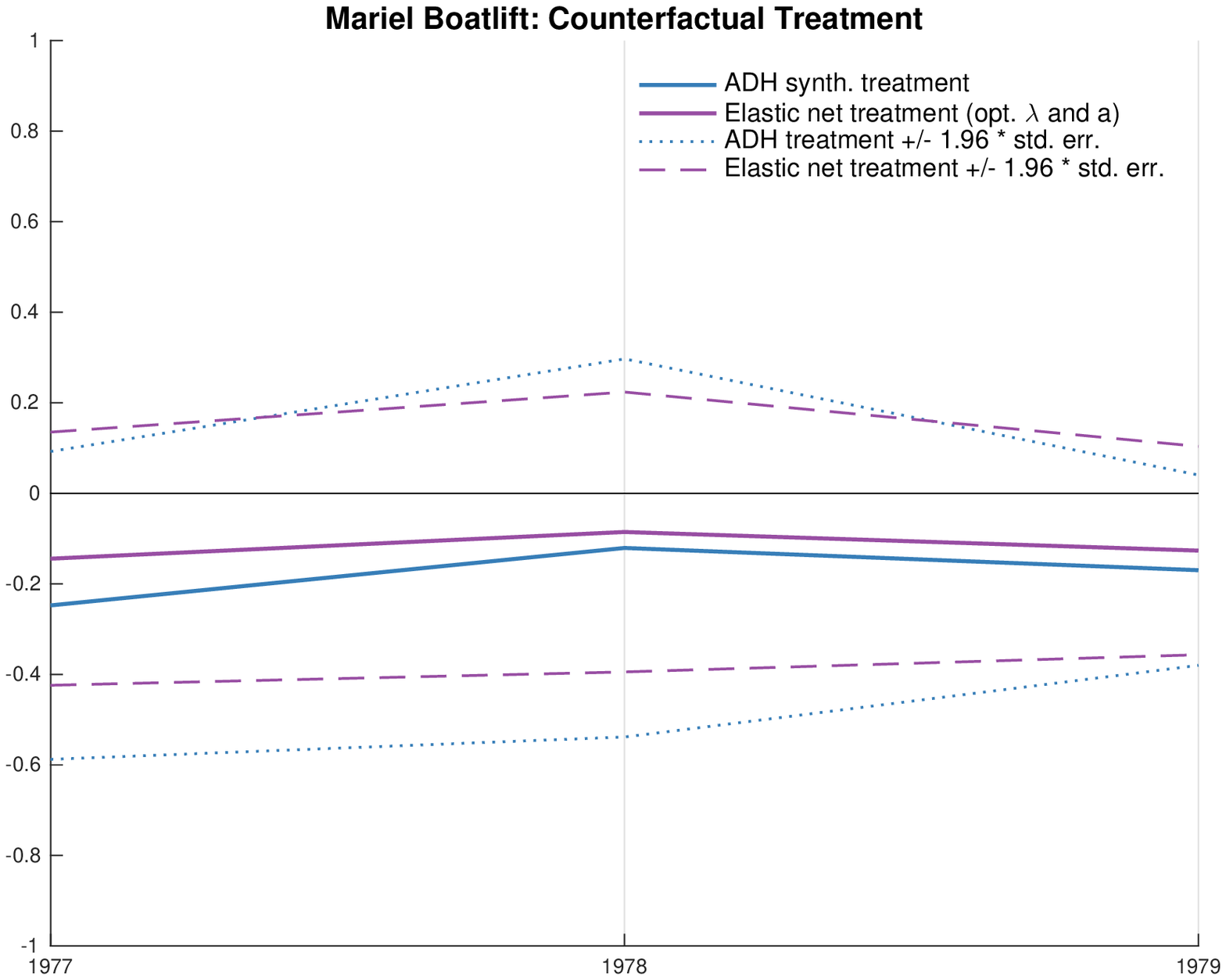}
  \end{subfigure}
\end{figure*}
\clearpage

\clearpage

\bibliographystyle{plainnat}
\bibliography{references}


\end{document}